\documentclass[a4paper]{article}

\usepackage{amsthm,amsmath}
\RequirePackage{hyperref}
\usepackage[utf8]{inputenc} %unicode support
\usepackage{graphicx}
\usepackage[margin=1in]{geometry}

\usepackage{units}
\usepackage{booktabs}
\usepackage{color}
\usepackage{cite} % for multi cites
\usepackage{microtype}
\usepackage{fancyhdr}

\fancypagestyle{firststyle}
{
   \fancyhf{}
   
   \lfoot{\small \textit{Preprint submitted to Advanced Modeling and Simulation in Engineering Sciences}}
}

\renewcommand{\vec}[1]{\boldsymbol{#1}}
\newcommand{\dd}{\mathrm{d}}

\newcommand{\pfrac}[2]{\frac{\partial #1}{\partial #2}}
\newtheorem*{remark}{Remark}

\title{On phase change and latent heat models in metal additive manufacturing process simulation}
\author{Sebastian D.~Proell, Wolfgang A.~Wall, Christoph Meier\thanks{
    Corresponding author: \textit{E-mail address}: \nolinkurl{meier@lnm.mw.tum.de}} \vspace{0.5em}\\
\small\textit{Technical University of Munich, Institute for Computational Mechanics,}\vspace{-0.3em}\\ \small\textit{Boltzmannstr.~15, 85748 Garching b.~München, Germany}}
\date{April 2020}
\begin{document}

%%% Start of article front matter
%\begin{frontmatter}

%\begin{fmbox}
%\dochead{Research}
\maketitle
\thispagestyle{firststyle}

\begin{abstract} % abstract
This work proposes an extension of phase change and latent heat models for the simulation of metal powder bed fusion additive manufacturing processes on the macroscale and compares different models with respect to accuracy and numerical efficiency. Specifically, a systematic formulation of phase fraction variables is proposed relying either on temperature- or enthalpy-based interpolation schemes. Moreover, two well-known schemes for the numerical treatment of latent heat, namely the apparent capacity and the so-called heat integration scheme, are critically reviewed and compared with respect to numerical efficiency and overall accuracy. Eventually, a novel variant of the heat integration scheme is proposed that allows to directly control efficiency and accuracy by means of a user-defined tolerance. Depending on the chosen tolerance, it is shown that this novel approach offers increased numerical efficiency for a given level of accuracy or improved accuracy for a given level of numerical efficiency as compared to the apparent capacity and the original heat integration scheme. The investigation and comparison of all considered schemes is based on a series of numerical test cases that are representative for application scenarios in metal powder bed fusion additive manufacturing.
\end{abstract}

\section{Introduction}
\label{intro}
Additive manufacturing (AM) is widely considered to be a key technology for future advances in engineering. AM offers highest flexibility in part design while still achieving the mechanical properties required for functional parts~\cite{Gibson2015}. In metal powder bed fusion additive manufacturing (PBFAM) multiple layers of metal powder are successively molten at selected positions, which eventually form the cross-sections of the final part after solidification. Energy is commonly deposited by a laser or electron beam giving rise to the respective names selective laser melting (SLM) and electron beam melting (EBM). These processes come with very challenging thermophysical phenomena on multiple length and time scales \cite{Herzog2016}. Accordingly, existing modeling approaches can be classified with respect to the resolved length scales: Macroscale approaches commonly aim at determining spatial distributions of physical fields such as temperature, residual stresses or dimensional warping on the scale of the design part. Mesoscale approaches resolve the length scale of individual powder particles on domains smaller than one powder layer to either study the melt pool thermo-fluid dynamics during the melting process \cite{Khairallah2014,Korner2011,Wessels2018,Markl2015,Russell2018,Yan2018} or the cohesive powder flow during the previous powder recoating process \cite{Herbold2015,Meier2019,Meier2019a}. Last, microscale approaches predict the evolution of the metallurgical microstructure during solidification \cite{Zhang2013,Gong2015,Rai2016,Salsi2018,Lindgren2016}. A broad overview of state-of-the-art modeling approaches on these different length scales can be found in \cite{Meier2017}. The present article focuses on the development of a thermal macroscale model for metal PBFAM processes.

Macroscale PBFAM models typically treat the powder phase as a homogenized continuum described via spatially averaged thermal and mechanical properties, without resolving individual powder particles. Also, the complex fluid dynamics within the melt pool is typically not explicitly resolved. Instead, a pure thermo-(solid-)mechanical problem is solved, usually based on a Lagrangian description and a spatial finite element discretization, with specific temperature- and phase-dependent thermal and mechanical constitutive parameters for the (homogenized) powder phase, the melt phase and the solidified phase. On the one hand, from a modeling point of view, such a procedure considerably simplifies the coupling of the different phase domains. On the other hand, this approach seems to be well-justified for certain simulations and quantities of interest since the mechanical forces transferred from powder and molten phase onto the solid phase are often negligible in good approximation.

In their works \cite{Gusarov2005,Gusarov2008,Gusarov2009}, Gusarov et al.\ proposed a model for powder bed laser absorption, which has been incorporated in many existing macroscale modeling approaches. For example, by using this absorption model, \cite{Hodge2014,Hodge2016} proposed a thermo-mechanical finite element (FE) model accounting for temperature- and phase-dependent thermal and mechanical constitutive behavior. Further developments in this field consider e.g. the accuracy of the physical model by adding additional physical effects such as residual stress relaxation \cite{Denlinger2015}, improved models for temperature and phase-dependent thermal conductivity of the powder \cite{Cervera1999,Childs2005} and melt \cite{Shen2012} phase, anisotropic conductivity \cite{Kollmannsberger2019}, phase-dependent laser absorptivity \cite{Roy2018}, thermodynamically consistent constitutive modeling based on phase energies  \cite{Bartel2018}, or by explicitly modeling the melt pool fluid dynamics \cite{Jamshidinia2013}, an approach inspired by similar schemes in the context of laser and electron beam welding \cite{Chang2015,Geiger2009,Rai2009}. Another important aspect for macroscale PBFAM models is computational efficiency, which has been addressed, among others, by applying dynamic mesh adaptivity schemes \cite{Kollmannsberger2017,Riedlbauer2017,Denlinger2014,Zhang2018a}, code parallelization and load balancing techniques~\cite{Neiva2019} as well as process layer agglomeration approaches
\cite{Zaeh2010,Hodge2016,Zhang2018a}.

The present work addresses two important aspects of macroscale PBFAM models, namely the modeling of phase change and latent heat effects.
Concerning the first aspect, we propose phase fraction variables which allow to formulate temperature- and phase-dependent material parameters in phase transition regions by consistent interpolation of the single phase parameters. While the definition of phase fraction variables is often somehow hidden in existing works, the present contribution defines these phase fraction variables in a transparent and systematic manner. Moreover, as basis of these phase fraction variables, different interpolation strategies, e.g. temperature-based or enthalpy-based interpolation, are discussed in detail. An alternative formulation of phase fractions based on energy minimization can e.g. be found in~\cite{Bartel2018}.

Concerning the modeling of latent heat effects, the two schemes that are most widely used in PBFAM models, namely the simple apparent (heat) capacity approach~\cite{Comini1974,DeMoraes2018,Morgan1978,Riedlbauer2017} and the more involved heat integration method~\cite{Rolph1982,Hodge2014,Oliveira2016}, will be investigated in the present work. It is known in the context of PBFAM process simulation that the specific choice of the latent heat model might considerably influence the overall efficiency of the numerical model~\cite{Hodge2014}. In the present work, the two aforementioned schemes, namely the apparent capacity approach and the heat integration method, are critically reviewed and compared with respect to numerical efficiency and overall accuracy. Eventually, a novel variant of the heat integration scheme is proposed that allows to directly control efficiency and accuracy by means of a user-defined tolerance. Depending on the chosen tolerance, it is shown that this novel approach offers increased numerical efficiency for a given level of accuracy or improved accuracy for a given level of numerical efficiency as compared to the apparent capacity and the original heat integration scheme.
One example where high accuracy, e.g. in the prediction of melt pool shape and thermal gradients, is essential is the modeling of microstructure evolution on the basis of temperature solutions provided by macroscale PBFAM models. While also the prediction of residual stresses can be considered as one of the major objectives of macroscale PBFAM process simulation, the present study intentionally focuses on purely thermal problems to isolate the effects of primary interest, namely the formulation and comparison of different phase change and latent heat models. For the same reason, local effects in the melt pool such as evaporation and fluid flow have been purposely neglected.

This article is structured as follows: Section~\ref{sec:thermal_model} presents the mathematical problem statement and summarizes the main model constituents in space- and time-continuous form. Specifically, Section~\ref{sec:laser_heat_source} introduces the heat source modeling of a laser beam. Next, in Section~\ref{sec:model_phase_latent} the phase change problem is introduced. It is shown how both mentioned methods for modeling latent heat, namely the apparent capacity and the heat integration scheme, can be derived from a Lagrange multiplier potential.  In Section~\ref{sec:temp_dep_params} temperature-history dependent material parameters are interpolated on the basis of properly defined phase fraction variables that allow to distinguish powder, solid and melt phase. The spatial and temporal discretization schemes as well as the general numerical solution procedure are outlined in Section~\ref{sec:numerics}. In Section~\ref{sec:numerics_details} the fully discretized version as well as algorithmic details of the heat integration method are discussed and eventually the novel tolerance-based variant of the heat integration scheme is proposed. Numerical experiments are presented in Section~\ref{sec:examples} with a focus on accuracy and efficiency of the considered methods. Finally, a summary of the present contribution and a brief outlook on future research work is given in Section~\ref{sec:conclusion}.

\section{General thermal model}

\label{sec:thermal_model}
For the purpose of this study it is sufficient to focus on the following transient purely thermal problem described by the heat equation for the temperature $T$ and appropriate boundary conditions. The problem statement in strong form is:
\begin{align}
\begin{aligned}
\label{eq:heat_equation}
c(T)\,\dot{T}+\nabla\cdot \vec{q} &= \hat{r}, &&\ \text{in}\ \Omega,\\
T &= \hat{T}, &&\ \text{on}\ \Gamma_T, \\
\vec{q}\cdot\vec{n} &= \hat{q}, &&\ \text{on}\ \Gamma_{\vec{q}}.
\end{aligned}
\end{align}
Temperatures $T$ are prescribed on the boundary part $\Gamma_T$  and  heat fluxes on $\Gamma_{\vec{q}}$. The heat flux $\vec{q}$ is specified by Fourier's law for isotropic material,
\begin{align}
\vec{q} = -k(T)\ \nabla T.
\end{align}
Material properties, namely (volumetric) heat capacity $c$ and conductivity $k$, may in general depend on the temperature $T$ but also on the phase $r$  (see Section~\ref{sec:temp_dep_params} for the definition of phase fractions and the interpolation of material parameters). In this contribution, spatial discretization will be based on the finite element method (see Section~\ref{sec:numerics}), i.e.~\eqref{eq:heat_equation} has to be transferred into its weak form via multiplication with a test function $\delta T$ and integration by parts, viz.
\begin{align}
\label{eq:heat_equation_weak}
\int_\Omega \delta T\,c(T)\dot{T}\, \dd\Omega - \int_\Omega \nabla\delta T\cdot\vec{q}\,\dd\Omega + \int_{\Gamma_{\vec{q}}}\delta T\,\hat{q}\,\dd\Gamma-\int_\Omega\delta T\, \hat{r}\,\dd\Omega = 0,
\end{align}
where the boundary conditions have already been inserted. By introducing the trial space $\mathcal{V} \!=\! \{T \, | \, T \!\in\! \mathcal{H}^1, T\!=\!\hat{T} \, \text{on} \, \Gamma_T \}$ as well as the weighting space \mbox{$\mathcal{W} \!=\! \{\delta T \, | \, \delta T \!\in\! \mathcal{H}^1, \delta T\!=\!0 \, \text{on} \, \Gamma_T \}$}, where $\mathcal{H}^1$ denotes the Sobolev space of functions with square-integrable first derivatives,~\eqref{eq:heat_equation_weak} is equivalent to the strong form~\eqref{eq:heat_equation}.
%%%%%%%%%%%%%%%%%%%%%%%%%%%%%%%%%%%%%%%%%%%%%%%%%%%%%%%%%%%%%%
\subsection{Modeling of the laser heat source}
\label{sec:laser_heat_source}
%%%%%%%%%%%%%%%%%%%%%%%%%%%%%%%%%%%%%%%%%%%%%%%%%%%%%%%%%%%%%%
The source term $\hat{r}$ represents the energy deposited by the laser beam as a volumetric heat source based on the model for radiative and conductive heat transfer in powder beds by Gusarov et al.~\cite{Gusarov2009}. In the following, a summary of the main model constituents is given. Let a powder layer be distributed in the $xy$-plane, where the powder material extends in positive z-direction from the powder layer surface at $z=0$ up to the layer thickness $L$ at $z=L$. A laser beam of nominal power $W$ and size $R$ is applied normal to this plane. The laser beam as well as the local coordinate system move in $x$-direction with a velocity $v$. The source term is then given in this local coordinate system relative to the laser beam center by
\begin{align}
\label{eq:mm_gusarov_laser_model}
\hat{r}(r_h,z) = -\beta_h Q_0 \pfrac{q}{\xi'},
\end{align}
where $r_h$ is the distance in the $xy$-plane from the laser beam center and $\beta_h$ is the extinction coefficient. The nominal power density $Q_0$ is radially distributed around the laser beam center as
\begin{align}
\label{eq:mm_gusarov_nominal_power_density_Q_0}
Q_0 = \begin{cases}
\frac{3W_e}{\pi R^2}\left(1-\frac{r_h}{R}\right)^2 \left(1+\frac{r_h}{R}\right)^2,\ &0 < r_h < R \\
0, &\text{otherwise}
\end{cases}.
\end{align}
The nominal laser power $W$ has been averaged and reduced to an effective power $W_e$ to account for various losses. Thus, \eqref{eq:mm_gusarov_nominal_power_density_Q_0} describes the spatial distribution in $x$- and $y$-direction. The normalized power density $q$ is given in terms of the dimensionless coordinate $\xi' = \beta_h z$ as
\begin{align}
\label{eq:mm_gusarov_normalized_density}
q &= \frac{\rho_h a}{(4\rho_h-3)D}\nonumber\\
&\left\{e^{-\lambda}(1-\rho_h^2)\left[e^{-2a\xi'}(1-a) + e^{2a\xi'}(a + 1)\right]\right.\nonumber\\
&-\left[e^{2a(\xi'-\lambda)}(1-a - \rho_h(a + 1))\right.\nonumber\\
&+\left.\left. e^{2a(\lambda - \xi')}(a + \rho_h(a - 1) + 1)\right](\rho_h e^{-2\lambda} + 3)\right\}\nonumber\\
&-\frac{3(1-\rho_h)(e^{-\xi'} - \rho_h e^{\xi'-2\lambda}
)}{4\rho_h - 3},
\end{align}
with hemispherical reflectivity $\rho_h$, constant $a=\sqrt{1-\rho_h}$, optical thickness $\lambda = \beta_h L$ and the constant
\begin{align}
D = &(1-a)\left[1-a-\rho_h(1+a)\right]e^{-2a\lambda}\nonumber\\
&-(1-a)\left[1+a-\rho_h(1-a)\right]e^{2a\lambda}.
\end{align}
Finally, the derivative of \eqref{eq:mm_gusarov_normalized_density} with respect to $\xi'$ as required in~\eqref{eq:mm_gusarov_laser_model} reads
\begin{align}
\label{eq:mm_gusarov_normalized_density_deriv}
\pfrac{q}{\xi'} &= \frac{(3-3\rho_h)(e^{-\xi'} + \rho_h e^{\xi'-2\lambda}
)}{4\rho_h - 3}\nonumber\\
&+\frac{2a^2\rho_h}{D(4\rho_h - 3)}\left\{e^{-\lambda}(1-\rho_h^2)\left[e^{-2a\xi'}(a - 1) + e^{2a\xi'}(a + 1)\right]\right.\nonumber\\
&+(\rho_h e^{-2\lambda} + 3)\nonumber\\
&\times\left.\left[e^{-2a(\lambda - \xi')}(a + \rho_h(a + 1) - 1) + e^{2a(\lambda - \xi')}(a + \rho_h(a - 1) + 1)\right]\right\}.
\end{align}
Apart from the optical properties,~\eqref{eq:mm_gusarov_normalized_density_deriv} only depends on the $z$-coordinate $z = \xi' / \beta_h $.
%%%%%%%%%%%%%%%%%%%%%%%%%%%%%%%%%%%%%%%%%%%%%%%%%%%%%%%%%%%%%%%%%%%%%%%%%%%%%%%%%%
\subsection{Modeling of phase change and latent heat} \label{sec:model_phase_latent}
\begin{figure}
    \centering
    \includegraphics[width=.9\linewidth]{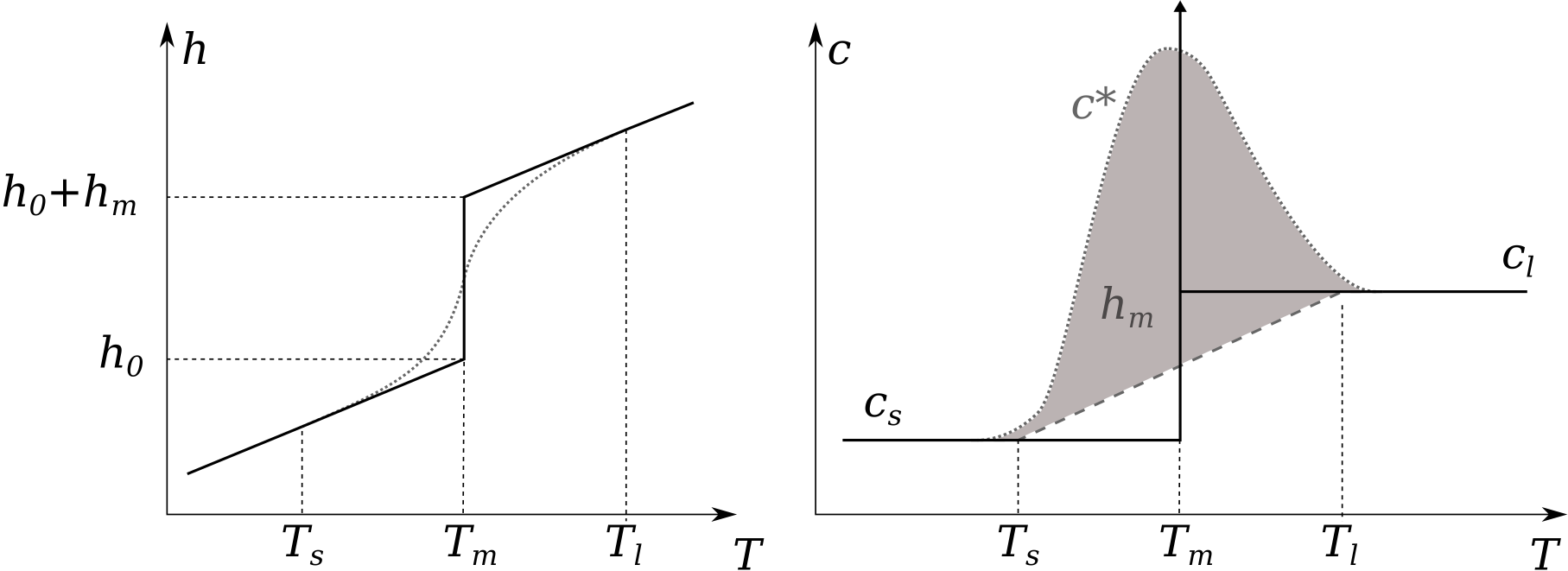}
    \caption{Left: enthalpy-temperature diagram for isothermal (solid line) and mushy (dotted line) phase change. Right: capacity-temperature diagram shows a singularity for isothermal phase change (solid line) at $T_m$. For the AC method the modified capacity $c^*$ includes the effects of latent heat $h_m$ within a regularized phase change interval $\left[T_s,T_l\right]$ similar to the mushy phase change type.}
    \label{fig:mm_apparent_heat_capacity}
    \end{figure}

So far we have not addressed the (crucial) phase change problem (i.e. melting and solidification), first from powder to molten and eventually from molten to solid phase. Consider a domain containing both solid and liquid phase separated by an interface $\Gamma_m$, which is defined by the isotherm $T = T_m$. Based on an energy balance at the interface, the following so-called Stefan-Neumann equation has to hold:
\begin{align}
\label{eq:mm_stefan_neumann_equation}
\underbrace{\vec{n}_{sl}\cdot\left(k_s\left.\pfrac{T}{\vec{x}}\right\vert_s-k_l\left.\pfrac{T}{\vec{x}}\right\vert_l\right)}_{\Delta q_m} =  h_m\vec{n}_{sl}\cdot\vec{v}_{sl}\quad\text{on}\ \Gamma_m,
\end{align}
where $\vec{n}_{sl}$ is the interface normal vector, $h_m$ the (volume-specific) latent heat of melting and the subscripts $(\cdot)_s$ and $(\cdot)_l$ denote quantities evaluated in the solid or liquid phase, respectively. The absorbed or released heat flux $\Delta q_m$ is proportional to the velocity $\vec{v}_{sl}$ of the evolving interface, in general leading to discontinuous heat fluxes across the phase interface. The free boundary condition~\eqref{eq:mm_stefan_neumann_equation} is especially suitable for sharp interface models in Eulerian description when combined with explicit interface tracking schemes, e.g.~via level set functions~\cite{Zhang2018a}, whose temporal evolution is defined by $\vec{v}_{sl}$. In this work, as typical for macroscale PBFAM models, the thermal problem is described in a Lagrangian manner in combination with a diffuse interface model, i.e. it is assumed that phase change takes place across an extended interface volume of finite thickness. Within this volume, the temperature is constrained until the melting enthalpy $h_m$ has been absorbed or released:
\begin{subequations}
\label{eq:interface_cond}
\begin{align}
\label{eq:interface_cond_absolute}
    T-T_m &= 0 &\text{if } h_0 \leq h \leq h_0+h_m\\
    \label{eq:interface_cond_rate}
    \dot{T} &= 0 &\text{if } h_0 \leq h \leq h_0+h_m,
\end{align}
\end{subequations}
where $h_0$ represents the enthalpy level at which melting starts. The left part of Fig.~\ref{fig:mm_apparent_heat_capacity} (solid line) illustrates this concept. When deriving the weak form~\eqref{eq:heat_equation_weak} in a variational manner, constraint equation~\eqref{eq:interface_cond_absolute} (or alternatively its rate form~\eqref{eq:interface_cond_rate}) can be enforced via the following Lagrange mutliplier potential
\begin{align}
    \Pi_m = \int_\Omega\lambda(T-T_m)\,\dd\Omega,
\end{align}
where $\lambda$ represents the Lagrange multiplier enforcing~\eqref{eq:interface_cond_absolute}. Its total variation yields
\begin{align}
\label{eq:lagr_mult_pot_variation}
    \delta \Pi_m = \int_\Omega\delta \lambda (T-T_m)\,\dd\Omega + \int_\Omega\delta T\, \lambda\,\dd\Omega.
\end{align}

The first term in~\eqref{eq:lagr_mult_pot_variation} represents the original constraint equation~\eqref{eq:interface_cond_absolute}, while the second term yields an additional contribution to the weak form~\eqref{eq:heat_equation_weak}:
\begin{align}
\label{eq:weak_form_with_lagr_mult}
    \int_\Omega \delta T\left(c(T)\dot{T} +\lambda\right)\, \dd\Omega - \int_\Omega \nabla\delta T\cdot\vec{q}\dd\Omega + \int_{\Gamma_{\vec{q}}}\delta T\,\hat{q}\,\dd\Gamma -\int_\Omega \delta T\, \hat{r}\,\dd\Omega= 0.
\end{align}
Since $\dot{T} = 0$ during phase change, the Lagrange multiplier equals the enthalpy rate
\begin{align}
\label{eq:lm_enthalpy_rate}
    \dot{h}(t) = \lambda(t) \quad \text{if} \quad h_0 \leq h \leq h_0+h_m,
\end{align}
which leads to the following expression for the enthalpy during phase change
\begin{align}
\label{eq:lm_enthalpy_integral}
    {h}(t) = h_0 + \int_{t(h_0)}^t \lambda(\tilde{t})\, \dd \tilde{t} \quad \text{if} \quad h_0 \leq h \leq h_0+h_m,
\end{align}
and eventually to an integral limiting condition for the Lagrange multiplier:
\begin{align}
\label{eq:accumnulated_metl_enthalpy_condition}
    0 \leq \int_{t(h_0)}^{t} \lambda(\tilde{t})\, \dd \tilde{t} \leq h_m.
\end{align}

Hu and Argyropoulos~\cite{Hu1996} review various methods to account for the latent heat associated with phase change. In the following two sections, the basics of two methods especially popular in the field of PBFAM modeling are briefly presented. In particular, we propose that these two schemes can be interpreted as different realizations of the constrained weak form~\eqref{eq:weak_form_with_lagr_mult}.

\begin{remark}
So far we have considered pure materials, i.e. isothermal phase change at a fixed melting temperature $T_m$. For alloys, phase change typically happens gradually, i.e. the latent heat is absorbed or released within a rather narrow temperature interval between solidus temperature $T_s$ and liquidus temperature $T_l$, in the following denoted as mushy phase change.
\end{remark}
\begin{remark} Throughout this work, the phase interface is implicitly defined by isotherms, a common choice in the context of PBFAM process simulation. Depending on the type of phase change (isothermal or mushy), one might take the isotherm at melting temperature (supplemented by a proper tolerance) or the isotherms at solidus and liquidus temperature to represent the interface.
\end{remark}

\subsubsection{Apparent capacity method}
\label{sec:app_capa}
One of the simplest approaches to capture the effects of latent heat is the so-called \textit{apparent capacity} (AC) method. The basic idea is to regularize the constraint~\eqref{eq:interface_cond} such that phase change takes place within a finite temperature interval of width $2d$ given by $T \in [T-d; T+d]$. Throughout this work, the temperature bounds confining this regularized phase change interval for isothermal phase change will be represented by the same variables $T_s=T-d$ and $T_l=T+d$ as the solidus and liquidus temperature for mushy phase change. This choice seems reasonable as it will turn out that the algorithmic treatment of both cases is identical. Considering now the weak from~\eqref{eq:weak_form_with_lagr_mult}, the apparent capacity method results from setting
\begin{align}
\label{eq:app_capa_lagr_mult}
    \lambda = c_m(T) \dot{T},
\end{align}
where the factor $c_m(T) \geq 0$ penalizes non-zero temperature rates $\dot{T} \neq 0$, i.e. violation of constraint~\eqref{eq:interface_cond_rate}. Thus, according to~\eqref{eq:app_capa_lagr_mult} the Lagrange multiplier is not considered as an independent primary variable and~\eqref{eq:interface_cond_rate} is not satisfied exactly anymore. Inserting~\eqref{eq:app_capa_lagr_mult} into~\eqref{eq:weak_form_with_lagr_mult} allows to identify a modified capacity,
\begin{align}
\label{eq:apparent_capacity}
    c^\ast(T) = c_m(T) +c(T),
\end{align}
justifying the name apparent capacity method. The choice of $c_m(T)$ is only restricted by the limiting condition~\eqref{eq:accumnulated_metl_enthalpy_condition}, which after a change of variable eventually reads
\begin{align}
\label{eq:apparent_capacity_condition}
    \int_{t(T_s)}^{t(T_l)} c_m(T) \dot{T} \, \dd t = \int_{T_s}^{T_l} c_m(T)\,\dd T \, \dot{=} \, h_m.
\end{align}

According to~\eqref{eq:apparent_capacity_condition}, small values of $c_m(T)$ require a large regularized phase change interval typically resulting in a more good-natured numerical algorithm at the cost of lower solution accuracy and vice versa. The original phase change constraint~\eqref{eq:interface_cond} (solid line) as well as the regularized phase change constraint based on the apparent capacity~\eqref{eq:apparent_capacity} (dashed line) are illustrated in Fig.~\ref{fig:mm_apparent_heat_capacity}. In this contribution a smoothed triangular distribution is chosen for $c_m$, as illustrated in the right part of Fig.~\ref{fig:mm_apparent_heat_capacity}.

As simple as the AC method may be, it suffers from one major drawback: An accurate representation of the phase change constraint, i.e. the choice of a small phase change interval $2d$ yields large values and steep gradients of the function $c_m(T)$, which are not only challenging for numerical solution schemes (e.g. nonlinear solvers, see Section~\ref{sec:numerics}) but also prone to large time integration errors. In this case, the absorbed or released enthalpy during phase change~\eqref{eq:apparent_capacity_condition}, and thus the overall energy balance of the phase change problem, is only captured with low accuracy.

\subsubsection{Heat integration method}
\label{sec:heatint}
Another popular and more advanced procedure is what Hu and Argyropoulos~\cite{Hu1996} call the \textit{heat integration} (HI) method. It has first been applied to a FE setting by Rolph and Bathe~\cite{Rolph1982} and is still used in more recent contributions \cite{Hodge2014,Oliveira2016}. Here, we present the basic concept by proposing an alternative interpretation of the heat integration method as an augmented Lagrange constraint enforcement scheme~\cite{laursen2002}. The full algorithmic details of the method in the spatially and temporally discretized problem setting will be presented in Section~\ref{sec:numerics_details}.

In a first step, the Lagrange multiplier in the weak form~\eqref{eq:heat_equation_weak} is replaced by an augmented Lagrange formulation for the constraint~\eqref{eq:interface_cond_absolute} of the form
\begin{align}
\label{eq:hi_augmented_lagrange}
    \lambda=\tilde{\lambda} + \epsilon (T-T_m)
\end{align}
where $\epsilon>0$ represents a penalty parameter. Comparable to an augmented Lagrange version based on the so-called Uzawa algorithm~\cite{laursen2002}, the new Lagrange multiplier $\tilde{\lambda}$ is not considered as an independent primary variable but rather as a history variable within an iterative solution procedure of the form $\tilde{\lambda}^i=\lambda^{i-1}$ leading to
\begin{align}
\label{eq:hi_uzawa_update}
\lambda^i=\lambda^{i-1} + \epsilon (T^{i}-T_m)  \quad \text{for} \quad i \geq 1 
\quad \text{with} \quad \lambda^{0}:=0,
\end{align}
where $i$ represents an iteration counter to be defined in Section~\ref{sec:numerics}. In the final algorithm, the enthalpy rate $\lambda^i$ at iteration $i$ has to fulfill the constraint equation~\eqref{eq:interface_cond_absolute} and the enthalpy inequality~\eqref{eq:accumnulated_metl_enthalpy_condition} up to a user-defined tolerance, which will eventually define a stopping criterion for the iterative procedure (see Section~\ref{sec:numerics_details_tolerance-based}).
\\

\begin{remark}
The HI scheme, which has originally been proposed in the time-discrete problem setting~\cite{Rolph1982}, can be recovered from~\eqref{eq:hi_uzawa_update} by choosing the penalty parameter according to $\epsilon =   c(T) / \Delta t$, where $\Delta t$ represents the time step size to be introduced in Section~\ref{sec:numerics}. Now, to motivate this specific choice and to illustrate the working principle of the HI scheme, consider the (theoretical) case of a material with constant capacity $c(T)=\text{const.}$ and vanishing conductivity $k(T)=0$. Then, the time-discrete version of heat equation~\eqref{eq:heat_equation} without any additional constraint, i.e. $i=0$ and $\lambda^0=0$ in~\eqref{eq:hi_uzawa_update}, yields the solution $c\Delta T^1 = \hat{r} \Delta t$ with $\Delta T^1 = \hat{r} \Delta t / c \neq 0$. If the considered material point is undergoing phase change in the current (and previous) time step, the solution can be expressed as $\Delta T^1 = T^1-T_m = \hat{r} \Delta t / c \neq 0$, which violates constraint~\eqref{eq:interface_cond_absolute}. Thus, an additional iteration $i=1$ with $\lambda^1=\epsilon (T^1-T_m)$ has to be conducted. For this example, the specific choice $\epsilon =   c / \Delta t$ allows to find the correct temperature
solution that is consistent with~\eqref{eq:interface_cond_absolute} already in the first iteration $i=1$:
\begin{align*}
\Delta T^2 = \hat{r}{\Delta t}/{c} - \lambda^1{\Delta t}/{c} = \hat{r}{\Delta t}/{c} - \underbrace{(T^1-T_m)}_{\hat{r} \Delta t / c} = 0
\end{align*}
Even though general scenarios with $c(T) \neq \text{const.}$ and $k(T) \neq 0$ will typically require a higher number of iterations, the choice $\epsilon =   c / \Delta t$ still seems reasonable and is well-established also in this case.
\end{remark}

\begin{remark}
The original definition of the Uzawa algorithm defines an additional fixed-point iteration wrapped around the iteration loop of the nonlinear solver (e.g. a Newton-Raphson scheme as discussed in Section~\ref{sec:numerics}) to perform the iterative update~\eqref{eq:hi_uzawa_update}. In contrast, the HI algorithm presented in Section~\ref{sec:numerics_details} will perform these updates directly during the iterations of the nonlinear solver without any additional ``outer" iteration loop.
\end{remark}

\subsection{Modeling of temperature- and phase-dependent parameters} \label{sec:temp_dep_params}
As common in macroscale PBFAM models, all three phases are modeled by varying material parameters of a solid material law. On the one hand, this procedure considerably simplifies the numerical schemes for coupling the different phases. On the other hand, this approach seems also to be justified for the thermo-mechanical problem since the mechanical forces transferred from powder and molten phase (with vanishing stiffness) onto the solid phase are negligible in good approximation for many questions. It is also important to note that the change from powder phase to molten phase is irreversible. In the following, we will consider the different types of phase changes in more detail.

\begin{figure}
    \centering
    \includegraphics[width=.6\linewidth]{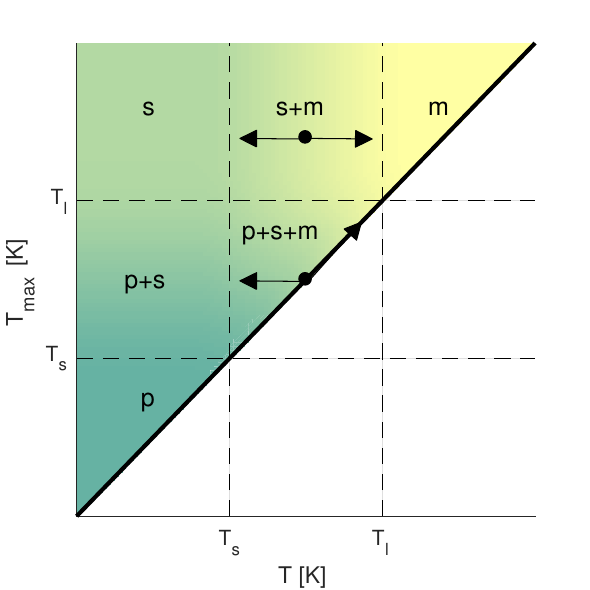}
    \caption{Temperature history diagram illustrating the two-dimensional nature of material parameter interpolation between powder (p), melt (m) and solid (s). Arrows indicate possible movement within the diagram.}
    \label{fig:mm_interpolation_phases}
    \end{figure}
Fig.~\ref{fig:mm_interpolation_phases} gives an illustration of the possible states of a material point. The current temperature $T$ is on the ordinate, while the abscissa shows the highest temperature ever reached, $T_\text{max}$. The different areas correspond to different mixtures of powder, melt and solid.
By definition there cannot be any possible state for $T>T_\text{max}$ and this area is blanked out.  If $T_\text{max}<T_s$, all material must still be in powder form (p). If $T_\text{max}>T_l$, there is no more powder, all material is consolidated and thus must be solid~(s) or molten (m).  The exact constitution of the two then depends on the current temperature.
The same reasoning applies to the current temperature. If $T>T_l$, all material must be molten. If $T<T_s$, material is a mixture of powder and solid, where the exact constitution is depending on $T_\text{max}$.
Perhaps the most interesting scenario is obtained when $T_s<T<T_\text{max}<T_l$. In this case some powder is still left and the consolidated phase consists of melt and solid. The arrows in Fig.~\ref{fig:mm_interpolation_phases} indicate the possible evolution of phases. Due to the definition of $T_\text{max}$ the only way to increase its value is an irreversible movement along the black diagonal line in Fig.~\ref{fig:mm_interpolation_phases}. For all other locations in the diagram a reversible horizontal movement is possible. With these considerations a temperature-based interpolation procedure for any material parameter can be derived.

\subsubsection{Temperature-based interpolation}
First, the focus is on the transition from (non-powder) solid to melt. The liquid fraction $g$ is introduced as

\begin{align}
\label{eq:mm_liquid_fraction}
g(T) = \begin{cases}
    0, & T < T_s\\
    \frac{T-T_s}{T_l-T_s}, &T_s \leq T \leq T_l\\
    1, & T > T_l
\end{cases}
\end{align}
If only solid and molten phase were present, any material parameter $f$ could be interpolated from the solid and melt values $f_s$ and $f_m$.
The history-dependent material behavior is captured by the fraction of consolidated material $r_c$ defined as
\begin{align}
\label{eq:mm_def_consolidated_fraction}
    r_c(t) = \begin{cases}
    1 & \text{, if } r_c(0) = 1 \text{ (i.e. initially consolidated)}\\
    \max\limits_{\tilde{t}\leq t}\, g(T(\tilde{t})) &\text{, if } r_c(0) = 0 \text{ (i.e. initially powder)}.\\
    \end{cases}
\end{align}
The time argument is explicitly stated to emphasize the history-dependence of $r_c(t)$. The consolidated fraction is initialized with a proper start value $r_c(0)$ (zero/one for locations initially covered with powder/consolidated material). For example, in a region that has initially been covered with powder, definition~\eqref{eq:mm_def_consolidated_fraction} equals the all-time maximum of the liquid fraction $g$ at this location which, according to~\eqref{eq:mm_liquid_fraction}, carries the same information as the maximum temperature $T_\text{max}$. In a region that has initially been covered with solid material, definition~\eqref{eq:mm_def_consolidated_fraction} equals one for all times since solid material can never transform back to powder.
The monotonously increasing fraction of consolidated material $r_c$ together with the liquid fraction $g$ allow to define volume fractions for powder, melt and solid phases according to:

\begin{align}
\label{eq:mm_powder_fraction}
r_p &= 1-r_c,\\
\label{eq:mm_molten_fraction}
r_m &= g,\\
\label{eq:mm_solid_fraction}
r_s &= r_c - g.
\end{align}
Their physical motivation is as follows: The powder fraction $r_p$ given in \eqref{eq:mm_powder_fraction} is by definition the complement of the consolidated fraction $r_c$. The molten fraction $r_m$ in \eqref{eq:mm_molten_fraction} is independent of the history and is always determined by \eqref{eq:mm_liquid_fraction}. The solid fraction $r_s$ defined in \eqref{eq:mm_solid_fraction} is the part of the consolidated fraction which is not molten. Note that definitions \eqref{eq:mm_powder_fraction}, \eqref{eq:mm_molten_fraction} and \eqref{eq:mm_solid_fraction} satisfy partition of unity and are thus suitable for interpolation.

Any material parameter $f$ can now be interpolated from the single phase values ${f_p, f_m, f_s}$ weighted by the corresponding fractions

\begin{align}
\label{eq:mm_phase_interpolation}
f(T) &= r_p(T)f_p + r_m(T)f_m + r_s(T)f_s.
\end{align}
For the special choice $f_p = f_s$, a two phase interpolation without history-dependent behavior would be recovered. The dependence on temperature is explicitly stated in \eqref{eq:mm_phase_interpolation} since this requires a consistent linearization to achieve robust convergence of the nonlinear solver.

\begin{figure}
    \centering
    \includegraphics[width=.8\linewidth]{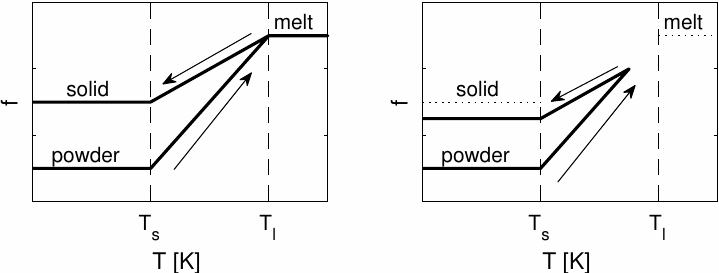}
    \caption{Evolution of a material parameter depending on temperature history. Left: material is completely molten and consolidated. Right: temperature history stays below liquidus temperature and thus a fraction of material stays powder.}
    \label{fig:interpolation_behavior}
    \end{figure}

\begin{figure}
    \centering
    \includegraphics[width=.6\linewidth]{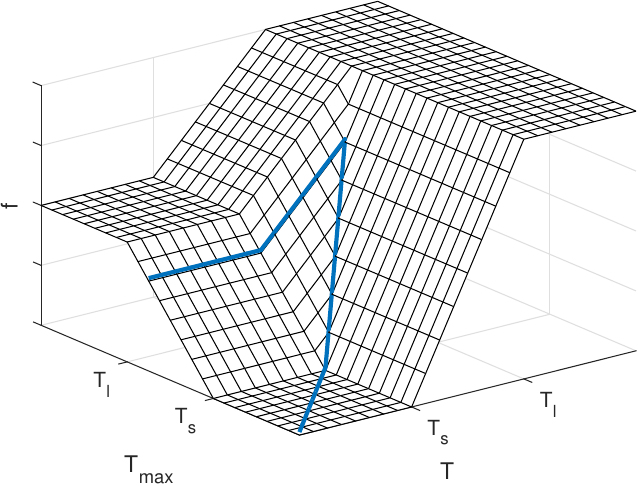}
    \caption{Three-dimensional visualization of parameter interpolation. Material parameters are continuous in $T$ and $T_\text{max}$ direction. The blue curve shows the evolution of the parameter when $T_\text{max}<T_s$.}
    \label{fig:interpolation_3d}
\end{figure}

Fig.~\ref{fig:interpolation_behavior} shows the evolution of an exemplary material parameter $f$ over the temperature history for the chosen liquid fraction definition \eqref{eq:mm_liquid_fraction}. In the left diagram powder melts completely and consequently the parameter $f$ takes on the value of a solid after cooling down below $T_s$. The right diagram shows partial melting: After cooling down below $T_s$ the parameter is a weighted average of the powder and solid value based on the consolidated fraction which is now smaller than 1. Fig.~\ref{fig:interpolation_3d} shows the same interpolation in a three-dimensional representation which makes the history-dependence on $T_\text{max}$ more explicit and shows that the parameter interpolation is also continuous over the history. This is important in the modeling of PBFAM processes since it ensures a continuous transition of material parameters between regions of molten or solidified material and regions that are still covered with unmolten powder.
Note that the definition of the liquid fraction~\eqref{eq:mm_liquid_fraction} determines the exact shape of the interpolating curve \eqref{eq:mm_phase_interpolation}. The kinks at $T_s$ and $T_l$ could also be smoothed out if this seems necessary for an improved numerical behavior (e.g. robust convergence of the nonlinear solver, see Section~\ref{sec:numerics}).

\begin{remark}
The definitions~\eqref{eq:mm_powder_fraction}, \eqref{eq:mm_molten_fraction} and \eqref{eq:mm_solid_fraction} imply that in a material mixture containing solid and powder the solid fraction always melts first, i.e. when increasing the temperature at a scenario $T_s<T<T_\text{max}<T_l$ the powder fraction does not increase before the previous maximum temperature $T_\text{max}$ is exceeded. With more history variables, also a different behavior could be realized (e.g. powder should melt first) but this is not deemed necessary in the context of macroscale PBFAM.
\end{remark}

\subsubsection{Enthalpy-based interpolation}
In the case of isothermal phase change (if not regularized by an AC scheme), the definition of the liquid fraction based on temperature as in~\eqref{eq:mm_liquid_fraction} is not directly applicable. A temperature-based parameter interpolation would still be possible based on a definition of solidus and liquidus temperature $T_s=T_m-d$ and $T_l=T_m+d$ as numerical regularization values as done in the AC scheme. However, a more consistent alternative approach utilizes the accumulated melt enthalpy $h(t)$ in~\eqref{eq:lm_enthalpy_integral} to define
\begin{align}
\label{eq:enth_based_liquid_fraction}
    g(t) = \begin{cases}
    0, & h = h_0\\
    \frac{h(t) - h_0}{h_m}, & h_0 < h < h_0 + h_m\\
    1, & h = h_0 + h_m
    \end{cases}
\end{align}
Essentially, the liquid fraction is defined as the ratio of already absorbed melt enthalpy to the latent heat required for phase change. All other parts of the interpolation~\eqref{eq:mm_phase_interpolation} remain unchanged. This liquid fraction will prove useful in combination with the HI method as further discussed in Section~\ref{sec:numerics_details}.

%%%%%%%%%%%%%%%%%%%%%%%%%%%%%%%%%%%%%%%%%%%%%%%%%%%%%%%%%%%%%%%%%%%%%%%%%%%%%%%%%%
\section{General numerical solution procedure}
\label{sec:numerics}
The weak form of the heat equation~\eqref{eq:heat_equation_weak} is discretized in space by the finite element method (FEM) following a Bubnov-Galerkin approach:
\begin{align}
\label{space_discr}
    T(\boldsymbol{x},t) = \sum_{j} N_j(\boldsymbol{x}) T_j(t), \quad \delta T(\boldsymbol{x}) =\sum_j  N_j (\boldsymbol{x}) \delta T_j,
\end{align}
where $\boldsymbol{x}$ is the spatial position, and $T_j(t)$ as well as $\delta T_j$ are the nodal temperatures and temperature variations. For time discretization a one-step theta time integration scheme is employed, which, for the model equation $\dot{\phi}=f(\phi,t)$, is defined as
\begin{align}
\label{temp_discr}
    \phi_{n+1} = \phi_n + \theta \Delta t f(\phi_{n+1},t_{n+1}) + (1-\theta) \Delta t f(\phi_n,t_n).
\end{align}
Here, $\Delta t:= t_{n+1} - t_{n}$ is the time step size and the subscript $(.)_n$ indicates a quantity that is evaluated at the discrete time step $t_{n}$. Together, the spatial and temporal discretization result in a fully discrete, nonlinear system of equations $\boldsymbol{R}(\boldsymbol{T}_{n+1}) = 0$, where $\boldsymbol{R}$ is the global residual vector and $\boldsymbol{T}_{n+1}$ the global vector of nodal temperatures at $t_{n+1}$. The equations are nonlinear due to the temperature-dependence of the heat capacity and thermal conductivity and due to the underlying phase change subproblem represented by (the non-smooth) constraint equation~\eqref{eq:interface_cond_absolute}. The system of nonlinear equations is linearized and solved iteratively with a Newton-Raphson scheme, which yields the following iterative update procedure
\begin{align}
\label{eq:newton-raphson}
    \boldsymbol{T}_{n+1}^{i+1} = \boldsymbol{T}_{n+1}^{i} + \Delta \boldsymbol{T}_{n+1}^{i+1} \quad \text{with} \quad \left.\pfrac{\boldsymbol{R}}{\boldsymbol{T}}\right\vert_{\boldsymbol{T}_{n+1}^i} \Delta \boldsymbol{T}_{n+1}^{i+1} = -\boldsymbol{R}(\boldsymbol{T}_{n+1}^i).
\end{align}
The following two convergence criteria $\vert\vert\boldsymbol{R}({T}_{n+1}^{i+1})\vert\vert_2 < \varepsilon_R$ and $\vert\vert \Delta\boldsymbol{T}_{n+1}^{i+1}\vert\vert_2 < \varepsilon_T$ are considered, i.e., for convergence both the norm of the residual and the iterative solution vector increment have to fall below given tolerances. All implementations and simulations presented in the next sections have been performed in the in-house research code BACI~\cite{Wall2019}, a parallel, multi-physics finite element framework.

\section{Discretization and algorithm of heat integration scheme}
\label{sec:numerics_details}
\textbf{Requirements and objective:} In Section~\ref{sec:examples}, it will turn out that the HI method typically describes the phase change problem more accurately (e.g. in terms of constraint~\eqref{eq:interface_cond_absolute}) as the AC scheme. Unfortunately, the HI method in its original form is known to typically result in a slow Newton-Raphson convergence~\cite{Hodge2014,Hu1996}, which can be traced back to residual manipulations in subsequent iteration steps $i$ on the basis of~\eqref{eq:hi_uzawa_update}, which are not accompanied by an associated consistent linearization contribution. In order to satisfy the constraint equation~\eqref{eq:interface_cond_absolute}, the original HI scheme typically leads to such manipulations, i.e. to changes of the Lagrange multiplier $\lambda^i$ in subsequent iterations, throughout the entire Newton-Raphson loop, which slows down convergence considerably. While the basics of this original HI scheme are presented in Section~\ref{sec:numerics_details_original}, we propose a novel realization of the HI scheme in Section~\ref{sec:numerics_details_tolerance-based}. This variant allows to control the accuracy of constraint enforcement in~\eqref{eq:interface_cond_absolute} via a user-defined tolerance. For practically relevant choices of this tolerance, $\lambda^i$ in~\eqref{eq:hi_uzawa_update} typically takes on a constant value after only a few iterations, thus allowing for quadratic convergence of the Newton-Raphson scheme afterwards.

 The basic idea of the HI method has been introduced in an abstract manner in Section~\ref{sec:heatint} for the case of isothermal phase change. Essentially, the method performs the update \eqref{eq:hi_uzawa_update} for the Lagrange multiplier $\lambda^i$ occurring in the discrete version of the weak form \eqref{eq:weak_form_with_lagr_mult} for each Newton-Raphson iteration $i$ at which constraint equation~\eqref{eq:interface_cond_absolute} is violated. It is important to note that the HI method in the spatially discretized setting enforces this constraint at element nodes (and not at integration points). For this purpose, the nodal volume $V_k$ is defined for each node $k$ as

\begin{align}
\label{eq:mm_heatint_nodalmass}
V_k = \int\limits_\Omega N_k\, \dd\Omega,
\end{align}

Taking advantage of the partition of unity $\sum_k N_k=1$, \eqref{eq:mm_heatint_nodalmass} implies that the sum of all the nodal volumes $V_k$ associated with an element indeed equals the volume occupied by this element. Considering the Lagrange multiplier term in~\eqref{eq:weak_form_with_lagr_mult}, the integral over $\Omega$ can then be approximated by a nodal evaluation of the integrand $\lambda$ times the nodal volume. Thus, the contribution to the residual entry $R_k$ yields:
\begin{align}
\label{eq:hi_nodal_H}
     R_k = \int_\Omega N_k \lambda \, \dd\Omega \approx  \lambda(\boldsymbol{x}_k) \int_\Omega N_k \, \dd\Omega = \lambda_k V_k =: \dot{H}_{k}
\end{align}
Since $\lambda_k$ is a volume-specific enthalpy rate, $\dot{H}_{k}$ can be interpreted as an absolute enthalpy rate. Similarly, the latent heat of melting associated with node~$k$ is

\begin{align}
\label{eq:heatint_total}
{H}_{m,k} = h_m V_k,
\end{align}

Employing~\eqref{eq:hi_uzawa_update} to express the Lagrange multiplier in~\eqref{eq:hi_nodal_H} at $t_{n+1}$ yields:
\begin{align}
\label{eq:hi_contribution_general}
    \!\!\!\!\!\!\!\! \dot{H}_{k,n+1}^i = V_k \lambda_{k,n+1}^i = V_k [\lambda_{k,n+1}^{i-1} + \epsilon (T_{k,n+1}^{i}-T_m)] =  \dot{H}_{k,n+1}^{i-1} + \frac{\Delta H_{k,n+1}^i}{\Delta t} ,\!\!\!\!
\end{align}
where the penalty parameter was chosen as $\epsilon = {c}/{\Delta t}$ (see Section~\ref{sec:heatint}) and hence
\begin{align}
\label{eq:hi_delta_H}
    \Delta H_{k,n+1}^i := c(T_{k,n+1}^{i}-T_m)V_k.
\end{align}
The iterative update rule~\eqref{eq:hi_contribution_general} together with~\eqref{eq:hi_delta_H} is the discrete counter part to~\eqref{eq:hi_uzawa_update}. Employing a backward Euler scheme for time integration of $\dot{H}_{k,n+1}^i$ in~\eqref{eq:hi_contribution_general} and assuming (without loss of generality) $h_0=0$ in~\eqref{eq:lm_enthalpy_integral}, the nodal enthalpy ${H}_{k,n+1}^i$ is
\begin{align}
\label{eq:hi_contribution_general2}
\begin{split}
    {H}_{k,n+1}^i & = {H}_{k,n} + \Delta t \dot{H}_{k,n+1}^i =  \underbrace{{H}_{k,n} +\Delta t \dot{H}_{k,n+1}^{i-1}}_{{H}_{k,n+1}^{i-1}} + \Delta H_{k,n+1}^i \\ &= {H}_{k,n+1}^{i-1} + \Delta H_{k,n+1}^i,
\end{split}
\end{align}
where ${H}_{k,n}$ represents the nodal enthalpy at the converged configuration of the last time step $t_n$. Thus, the iterative update rule~\eqref{eq:hi_contribution_general2} for the nodal enthalpy ${H}_{k,n+1}^i$ has the same form as the update rule~\eqref{eq:hi_contribution_general} for the nodal enthalpy rate $\dot{H}_{k,n+1}^i$ but with the different initial values ${H}_{k,n+1}^0={H}_{k,n}$ and $\dot{H}_{k,n+1}^0=0$ required in the first iteration $i=1$ of a time step. Also the enthalpy limiting condition~\eqref{eq:accumnulated_metl_enthalpy_condition} during phase change can be transferred to the discrete problem setting, which reads (for $h_0=0$):
\begin{align}
\label{eq:hi_contribution_limiting1}
    0 \leq {H}_{k,n+1}^i \leq  H_{m,k}.
\end{align}
Essentially, the working principle of the heat integration scheme is to add residuum contributions in~\eqref{eq:hi_nodal_H} defined by the iterative update scheme~\eqref{eq:hi_contribution_general} as long as the discrete limiting condition~\eqref{eq:hi_contribution_limiting1} is satisfied. The detailed algorithm of the original HI scheme as well as the required adaptions for the proposed tolerance-based HI scheme will be presented in the next two sections.

\begin{remark}
As noted by the authors of the original work~\cite{Rolph1982} and in accordance with~\eqref{eq:hi_nodal_H}, the HI method requires the consistent use of the nodal lumped capacity. Thus, when using this algorithm, the capacity matrix must enter the residual and Jacobian in the Newton-Raphson algorithm \eqref{eq:newton-raphson} in lumped form to obtain a robust scheme. Note also that no linearization contribution associated with the residual term in~\eqref{eq:hi_nodal_H} is considered in the context of the Newton-Raphson method.
\end{remark}

\subsection{Original heat integration scheme}
\label{sec:numerics_details_original}

We will first introduce our understanding of the original HI method~\cite{Rolph1982}. The specific notation was inspired by the formulation in~\cite{Oliveira2016}. So far, the considerations have been restricted to isothermal phase change but they can be extended to the treatment of mushy phase change. The only difference is the fact that during melting the temperature is not constrained to the constant value $T_m$ but rather to a gradually evolving intermediate temperature $T'$ between $\left[T_s,T_l\right]$. Thus, in the following, we will present the more general algorithm for mushy phase change and point out the relevant differences to the isothermal scenario.

Each node $k$ stores the nodal enthalpy ${H}_{k,n+1}^{i-1}$ as a history-variable. At the beginning of the simulation, it is initialized to zero for solid material. At the first Newton-Raphson iteration of a time step it is initialized with the converged value from the last time step, i.e. ${H}_{k,n+1}^{0}={H}_{k,n}$. Now, after each Newton-Raphson iteration $i$ in time step $n+1$ the following calculations are performed:
\begin{enumerate}
\item Skip node $k$ if

\begin{align}
\label{eq:heat_int_skip_orig}
&\left[T_{k,n+1}^i < T_s\ \text{and}\ T_{k,n} < T_s\right]\ \textbf{or}\nonumber\\
&\left[T_{k,n+1}^i > T_l\ \text{and}\ T_{k,n} > T_l\right]
\end{align}
which means it is not undergoing phase change. In this case the increment $\Delta H_{k,n+1}^{i}$ in~\eqref{eq:hi_contribution_general} is set equal to zero. Here, $T_{k,n}$ denotes the converged temperature value of node $k$ at the last time step $n$.
\item Else, for each node $k$ which is undergoing phase change
compute the increment

\begin{align}
\label{eq:heat_int_increment}
\Delta H_{k,n+1}^{i} = c'\left(T_{k,n+1}^{i} -T'\right)V_k.
\end{align}

%which is the lumped version of \eqref{eq:mm_heatint_incr_source}.
Here, $T'$ is an intermediate temperature given by

\begin{align}
\label{eq:mm_heatint_intermediate_temp}
T' = T_s + \left\vert\frac{{H}_{k,n+1}^{i-1}}{H_{m,k}}\right\vert\, (T_l-T_s),
\end{align}

calculated from the amount of latent heat already absorbed (released) during melting (freezing). The modified capacity is computed as

\begin{align}
c' = \left({\frac{T_l-T_s}{h_m} + \frac{2}{c_s+c_l}}\right)^{-1},
\end{align}

where $c_s$ and $c_l$ are the values of heat capacity at $T_s$ and $T_l$, respectively. In case of isothermal phase change, the intermediate temperature simplifies to the melting temperature, i.e. $T' = T_m = T_s = T_l$, and the modified capacity simplifies to the averaged capacity at the melting point, i.e. $c' = \frac{1}{2}(c_s+c_l)$ such that~\eqref{eq:heat_int_increment} becomes identical to~\eqref{eq:hi_delta_H}.
\item Limit each increment $\Delta H_{k,n+1}^{i}$ such that the following condition is fulfilled:

\begin{align}
\label{eq:mm_heatint_limiting_cond}
0 \leq H_{k,n+1}^{i-1} + \Delta H_{k,n+1}^{i} \leq  H_{m,k}.
\end{align}

If condition~\eqref{eq:mm_heatint_limiting_cond} is violated, the increment $\Delta H_{k,n+1}^{i}$ is limited such that the respective bound in~\eqref{eq:mm_heatint_limiting_cond} is exactly met (e.g. $\Delta H_{k,n+1}^{i} = H_{m,k} - H_{k,n+1}^{i-1}$ if the right bound in~\eqref{eq:mm_heatint_limiting_cond} was exceeded). Afterwards update the nodal enthalpy:
\begin{align}
\label{eq:hi_contribution_general3}
    {H}_{k,n+1}^i = {H}_{k,n+1}^{i-1} + \Delta H_{k,n+1}^i \quad \text{with} \quad {H}_{k,n+1}^{0}={H}_{k,n}.
\end{align}

\item For each node $k$ where $\vert\Delta H_{k,n+1}^i\vert > 0$ reset the temperature
\begin{align}
\label{eq:mm_heatint_tempreset}
T_{k,n+1}^i  = T'.
\end{align}

\item Calculate the updated enthalpy rate according to~\eqref{eq:hi_contribution_general},
\begin{align}
\label{eq:mm_heatint_sourceterm}
    \dot{H}_{k,n+1}^i = \dot{H}_{k,n+1}^{i-1} + \frac{1}{\Delta t} \Delta H_{k,n+1}^i \quad \text{with} \quad \dot{H}_{k,n+1}^{0}=0,
\end{align}
and add the residual contribution~\eqref{eq:hi_nodal_H}. The start value $\dot{H}_{k,n+1}^0 = 0$ in the first iteration $i=1$ of each time step is the equivalent of $\lambda^0$ in~\eqref{eq:hi_uzawa_update}.
\end{enumerate}
It is emphasized that limiting condition~\eqref{eq:mm_heatint_limiting_cond} allows to use the same algorithm for both melting and solidification as discussed in the remark at the end of this section.

\subsection{Tolerance-based heat integration scheme}
\label{sec:numerics_details_tolerance-based}

As stated at the beginning of Section~\ref{sec:numerics_details}, the original HI method is known to result in a slow Newton-Raphson convergence due to the residual manipulations~\eqref{eq:mm_heatint_sourceterm}, which are not accompanied by an associated consistent linearization contribution and which would typically occur throughout the entire Newton-Raphson loop if no additional, tolerance-controlled abort criterion is utilized. To improve the algorithm, we propose to introduce a tolerance $\varepsilon_\text{tol}$ to control the accuracy of the phase change representation. Hence, we stop adding increments $\Delta H_{k,n+1}^i$ in the algorithm above when these are small compared to the latent heat of melting, i.e.,
\begin{align}
\label{eq:mm_relative_increment}
\left\vert\frac{ \Delta H_{k,n+1}^i}{{H}_{m,k}}\right\vert < \varepsilon_\text{tol}.
\end{align}

where $\varepsilon_\text{tol}<1$ is a relative tolerance describing the relative amount of latent heat which is not absorbed (released) after melting (solidification) is complete. Inserting \eqref{eq:heatint_total} and \eqref{eq:heat_int_increment} into \eqref{eq:mm_relative_increment} yields an alternative for step 1 in the algorithm above:
\begin{itemize}
\item [1$^\ast$] Skip node $k$ if

\begin{align}
\label{eq:heatint_skip_tolerance}
\vert T_{k,n+1}^i - T' \vert < \varepsilon_\text{tol}\frac{ h_m}{ c'}.
\end{align}

\end{itemize}
The new criterion provides a way to stop the HI algorithm when constraint~\eqref{eq:interface_cond_absolute} is satisfied with a certain accuracy. At first glance, the new criterion~\eqref{eq:heatint_skip_tolerance} seems to significantly change the outcome of the algorithm since \eqref{eq:heat_int_increment} will be evaluated for nodes that are far away from a phase change. However, it can easily be verified that the signed limiting condition~\eqref{eq:mm_heatint_limiting_cond} will automatically skip all nodes that are not meant to undergo phase change on the basis of their current temperature value (see also the discussion in the remark below).

\begin{remark}
Consider a node that is heated up and the material is undergoing a melting process. Initially, the node is in solid state, i.e. $H_{k,n+1}^{i-1}=0$, with a temperature below the solidus temperature, i.e. $T_{k,n+1}^i < T_s$. The calculated increment \eqref{eq:heat_int_increment} would be negative. Since there is no phase change yet, no increment should be added. Indeed, limiting according to \eqref{eq:mm_heatint_limiting_cond} will not allow a negative increment with a zero history and the increment is set to zero. When the temperature rises above the solidus temperature $T_s$, however, the increments become positive and will be considered according to~\eqref{eq:mm_heatint_limiting_cond}. Positive increments are added to the the nodal latent heat $H_{k}$ until it reaches the allowed maximum $H_{m,k}$. %which by definition~\eqref{eq:heatint_total} is also negative.
Then the material is fully molten.

Next, consider the inverse process, i.e. cooling of a node with material initially in the molten state. If material is molten, then $H_{k,n+1}^{i-1} = H_{m,k}$. Looking at the limiting condition~\eqref{eq:mm_heatint_limiting_cond} shows that only negative increments are allowed, when material is molten. As expected, negative increments~\eqref{eq:heat_int_increment} are obtained, and the solidification process is initiated, as soon as temperature drops below the liquidus temperature $T_l$. The negative increments are added to $H_{k}$ until it reaches a value of zero. Then the material has returned to the solid state.
The phase change as it was just described is fully reversible.
\end{remark}

\begin{remark}
Now that the details of the HI scheme have been introduced, the enthalpy-based liquid fraction~\eqref{eq:enth_based_liquid_fraction} used for parameter interpolation in Section~\ref{sec:temp_dep_params} can be further specified. For node $k$ it reads
\begin{align}
\label{eq:mm_heatint_enthaply_liquid_fraction}
g_k = \frac{H_{k,n+1}^i}{H_{m,k}}.
\end{align}
In the numerical examples of the following section, this liquid fraction based on latent heat will be employed in combination with the isothermal HI scheme. All latent heat schemes employing a finite phase change interval $[T_s;T_l]$, will use the temperature-based parameter interpolation with liquid fraction~\eqref{eq:mm_liquid_fraction}. 
\end{remark}

%%%%%%%%%%%%%%%%%%%%%%%%%%%%%%%%%%%%%%%%%%%%%%%%%%%%%%%%%%%%%%%%%%%%%%%%%%%%%%%%%%
\section{Numerical results}
\label{sec:examples}
\subsection{Solidification front in a 1D slab}
\label{sec:results_solidifcation_front}
\begin{figure}
    \centering
    \includegraphics[width=.8\linewidth]{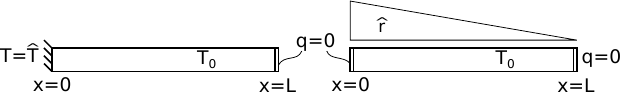}
    \caption{Geometry, thermal loads, boundary and initial conditions for solidification front example (left) and melting volume example (right).}
    \label{fig:geometry_solidification_melting}
    \end{figure}

To validate the implementation, first a series of numerical experiments are conducted on a one dimensional domain for which an analytic solution is available~\cite{Hu1996}.
This example has already been used  to show the validity of methods for capturing latent heat~\cite{Comini1974}. A pseudo one-dimensional slab (material properties of ice / water) with length $L=\unit[1]{m}$ is subject to a fixed temperature $\hat{T}=\unit[253]{K}$ on its left edge at $x=0$. The initial temperature in the whole slab is $T_0=\unit[283]{K}$. The left part of Fig.~\ref{fig:geometry_solidification_melting} illustrates this scenario.
The interface separating frozen and liquid water will slowly travel from left to right. Material parameters for water are given in Table~\ref{tab:mm_semi-inf_slab_params}, they are taken directly from \cite{Comini1974}. The problem is discretized with 25, 50 or 100 linear finite elements in space and three different fixed step sizes $\Delta t \in \{\unit[200]{s},\unit[400]{s},\unit[800]{s}\}$ in time. Total simulation time is $t_f = \unit[72\cdot 10^3]{s}$. At this point in time the temperature on the right edge is still at the initial level and the analytic solution (which is calculated on a semi-infinite domain) remains valid.

    \begin{table}
        \centering
        \caption{Material parameters of water \cite{Comini1974} for Sections~\ref{sec:results_solidifcation_front} and~\ref{sec:results_melting_vol}.}
        \label{tab:mm_semi-inf_slab_params}
        \begin{tabular}{llll}
        \toprule%\hline\noalign{\smallskip}
        Parameter & Description & Value & Unit\\
        \midrule%\noalign{\smallskip}\hline\noalign{\smallskip}
        $T_m$ & Phase change temperature & 273 & K\\
        $h_m$ & Volumetric latent heat & 338 & \unitfrac{MJ}{m$^3$}\\
        $c_s$ & Volumetric specific heat, solid & 1.762 & \unitfrac{MJ}{m$^3$K}\\
        $c_l$ & Volumetric specific heat, liquid & 4.226 & \unitfrac{MJ}{m$^3$K}\\
        $k_s$ & Thermal conductivity, solid & 2.22 & \unitfrac{W}{mK}\\
        $k_l$ & Thermal conductivity, liquid & 0.556 & \unitfrac{W}{mK}\\
        \bottomrule%\noalign{\smallskip}\hline
        \end{tabular}
        \end{table}

The introduced HI method will be used in four variants by distinguishing a) isothermal and mushy phase change as well as b) the original criterion~\eqref{eq:heat_int_skip_orig} and the novel tolerance-based criterion~\eqref{eq:heatint_skip_tolerance}.
In this example, phase change of water is isothermal and thus the isothermal HI methods with either original or tolerance-based criterion can be applied directly. For the AC method an artificial phase change interval is chosen with $T_s=\unit[270]{K}$ and $T_l=\unit[276]{K}$, i.e., $d=\unit[3]{K}$. The same interval is used to apply the original and tolerance-based mushy HI methods. Additionally, for tolerance-based HI the tolerance is chosen as $\varepsilon_\text{tol}=0.001$, i.e., up to 0.1\% of latent heat will be neglected.

\begin{figure}
    \centering
    \includegraphics[width=.6\linewidth]{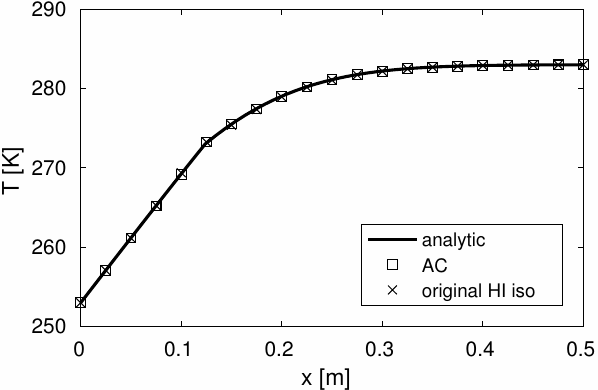}
    \caption{Melting front example: temperature profiles at $t_f$ of analytic solution compared to numerical results with AC and HI method. 100 elements, $\Delta t = \unit[200]{s}$.}
    \label{fig:compare_heatintegration_apparentcapa}
    \end{figure}

\begin{figure}
        \centering
        \includegraphics[width=.8\linewidth]{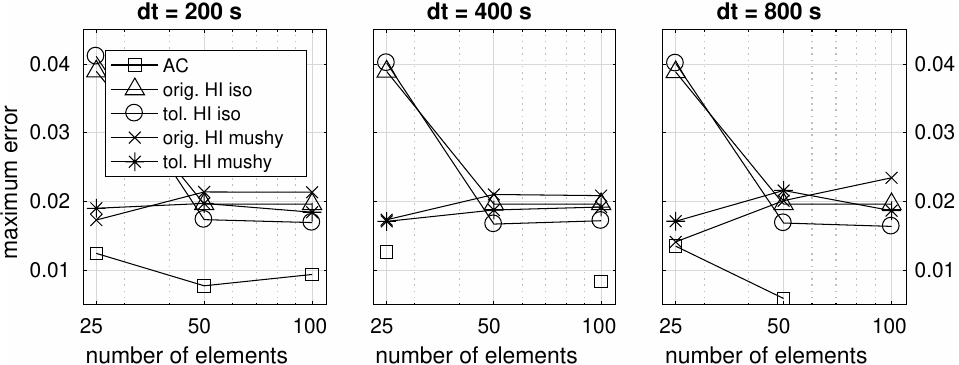}
        \caption{Melting front example: maximum error in temperature $\left\Vert\frac{T-T_\text{ref}}{T_0-\hat{T}}\right\Vert_\infty$ of different methods for latent heat depending on spatial and temporal discretization. Missing data points for the AC scheme indicate that no convergence was achieved for these parameter choices.}
        \label{fig:compare_maxerr_discretization}
        \end{figure}

All approaches yield results that agree very well with the analytic solution. Fig.~\ref{fig:compare_heatintegration_apparentcapa} shows the solutions obtained with AC and original isothermal HI method on the finest mesh as an example. The maximum errors in the numerical solutions provided by the different methods are shown for the three investigated meshes and time step sizes in Fig.~\ref{fig:compare_maxerr_discretization}. The maximum errors lie below 4\% for the coarsest mesh and around 2\% for the finer meshes, which is deemed accurate enough for the intended use case. Within the considered scope, the time step size seems to have only minor effect on the accuracy. All versions of HI schemes produce errors that are slightly higher compared to the error from the AC scheme. However, for the large time step sizes $\Delta t = \unit[400]{s}$ and $\Delta t = \unit[800]{s}$ the AC method does not always converge.

\begin{figure}
    \centering
    \includegraphics[width=0.8\textwidth]{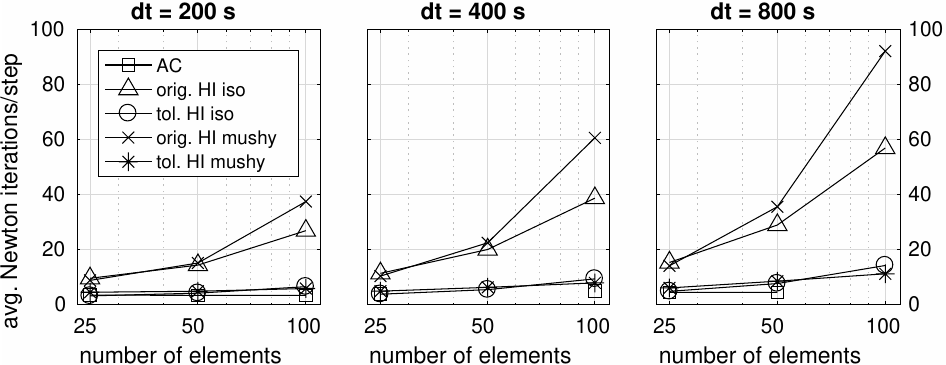}
    \caption{Melting front example: average number of iterations of different methods for latent heat depending on spatial and temporal discretization. Missing data points for the AC scheme indicate that no convergence was achieved for these parameter choices.}
    \label{fig:compare_iterations_discretization}
    \end{figure}

The real difference between the methods comes to light when numerical efficiency is investigated in terms of Newton iterations needed per time step. We choose to analyze Newton iterations as a measure for the efficiency of the proposed methods instead of computational time, which typically leads to a stronger dependency on the specific code implementation. Still, the resulting computation time, which is usually of practical interest, scales directly with the number of Newton iterations. Fig.~\ref{fig:compare_iterations_discretization} shows a strong dependence of the original HI method~\cite{Rolph1982} on the spatial discretization with heavily increased iterations per time step in case of finer spatial resolution. On top of that a larger time step leads to a further increase in iterations. Both isothermal and mushy version of the original method suffer from this effect. Hodge et al.~\cite{Hodge2014} mention that small time steps had to be used because of the original HI method. However, our proposed tolerance-based method does not only require significantly less iterations per step in every scenario but is also less sensitive to spatial and temporal discretization. This seems beneficial when moving to simulation of PBFAM processes on a part-scale.

\subsection{Melting volume in a 1D slab}

\label{sec:results_melting_vol}
We investigate the same variants on a slightly modified second example (Fig.~\ref{fig:geometry_solidification_melting}, right). The Dirichlet condition on the left boundary is dropped and all faces are assumed to be insulating. Instead a spatially varying source term $\hat{r} = 20,000(1-x)\unitfrac{W}{m^3}$ is applied to the whole slab of water which is initially frozen at $T_0=\unit[263]{K}$. Material parameters for solid and liquid water are again given in Tab.~\ref{tab:mm_semi-inf_slab_params}.  In contrast to the previous example, melting  will not take place at a single node representing the phase interface. Instead a whole volume can be in phase transition (i.e. melting). The same spatial and temporal discretizations from before are used, total simulation time is $t_f = \unit[20\cdot 10^3]{s}$.

\begin{figure}
    \centering
    \includegraphics[width=.9\linewidth]{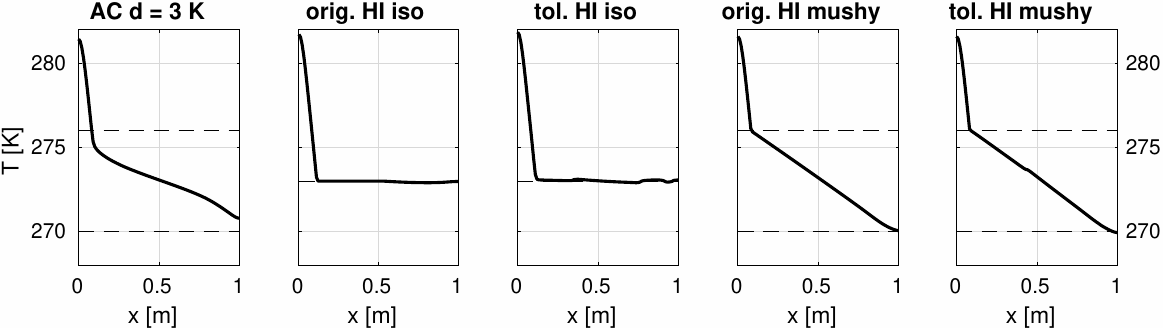}
    \caption{Melting volume example: temperature profiles at $t_f$ obtained with five different methods for latent heat. 100 elements, $\Delta t = \unit[200]{s}$.}
    \label{fig:meltvol_temp_ele100_dt200}
    \end{figure}

Again, the AC method and the four variants of HI are applied with the settings from above. No analytic solution is available for this scenario. Fig.~\ref{fig:meltvol_temp_ele100_dt200} shows the obtained temperature profiles along the one dimensional slab for the fine discretization with 100 elements and a step size of $\unit[200]{s}$. Obviously, AC and mushy HI methods will not keep temperatures fixed at the melting temperature of \unit[273]{K} during phase change. When one is concerned about exact representation of isothermal phase change, only the original and tolerance-based version of isothermal HI are accurate enough, although some oscillation around the melting temperature is observed. Looking at the final temperatures on the left edge, when melting is already finished, reveals that the latent heat of melting still is captured with good accuracy by all methods, and the predicted temperatures agree well. Given the large temperature range prevalent in the targeted application (PBFAM simulation on part-scale), a highly accurate representation of temperature profiles around the melting point is not of highest practical importance.

\begin{figure}
    \centering
    \includegraphics[width=.7\linewidth]{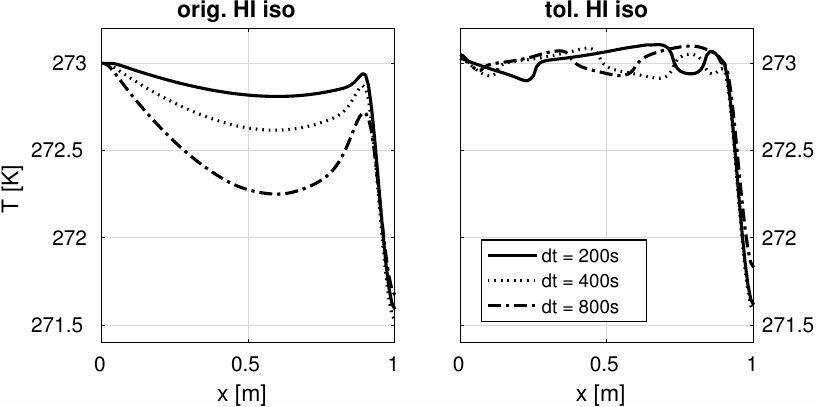}
    \caption{Melting volume example: temperature profiles at $0.5\,{t_f}$ obtained from the original and the proposed tolerance-based HI scheme (100 elements, time step varying).}
    \label{fig:meltvol_heatint_rb_tol}
    \end{figure}

\begin{figure}
    \centering
    \includegraphics[width=.7\linewidth]{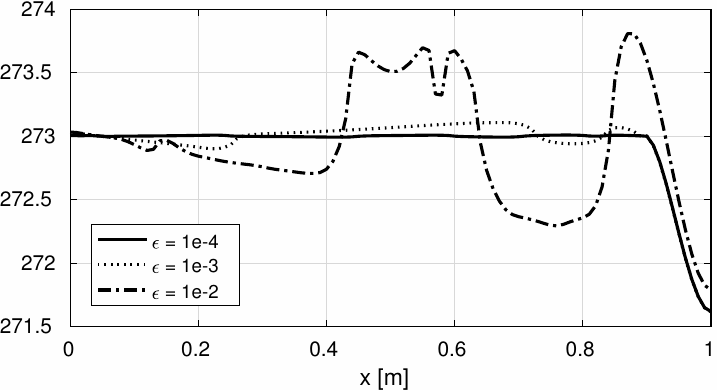}
    \caption{Melting volume example: temperature profiles at $0.5\,{t_f}$ obtained from the proposed tolerance-based HI scheme for different tolerances (100 elements, $\Delta t=\unit[200]{s}$).}
    \label{fig:meltvol_heatint_tol_var}
    \end{figure}

A short look on the accuracy of isothermal HI schemes is taken in this paragraph.
Solutions with the HI scheme can become quite inaccurate around the melting temperature especially when larger time steps are used. Fig.~\ref{fig:meltvol_heatint_rb_tol} shows such solutions of the original scheme at $t=0.5t_f$ for different step sizes and compares it to our proposed tolerance-based method. Both methods cannot exactly enforce isothermal phase change which would be characterized by a horizontal plateau region at $T_m$. The original method's criterion \eqref{eq:heat_int_skip_orig} for determining nodes undergoing phase change proves to be ill-suited for this scenario. Larger time steps lead to larger undershoots in temperature. The fluctuations around melting temperature obtained with the proposed tolerance-based HI scheme on the contrary are independent of step size and only controlled by the tolerance $\varepsilon_\text{tol}$. The temperature profiles resulting from three different tolerances (0.01, 0.001 and 0.0001) can be seen in Fig.~\ref{fig:meltvol_heatint_tol_var}.

    \begin{figure}
        \centering
        \includegraphics[width=.8\linewidth]{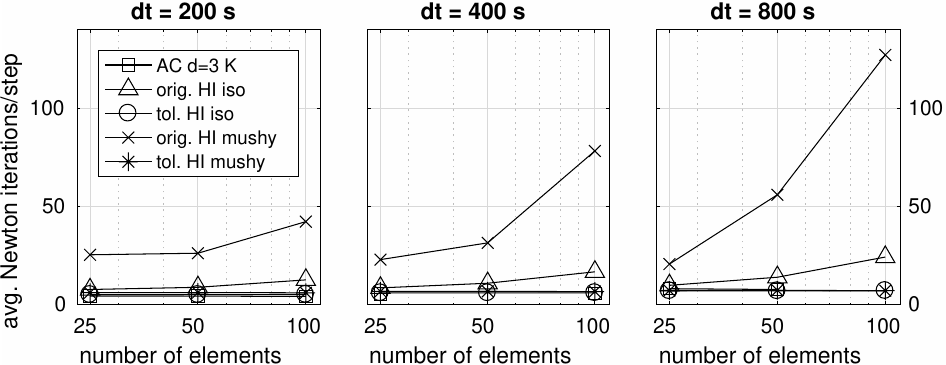}
        \caption{Melting volume example: average number of iterations of all investigated methods for latent heat depending on spatial and temporal discretization.}
        \label{fig:compare_iterations_discretization_meltvol}
        \end{figure}

\begin{figure}
    \centering
    \includegraphics[width=.8\linewidth]{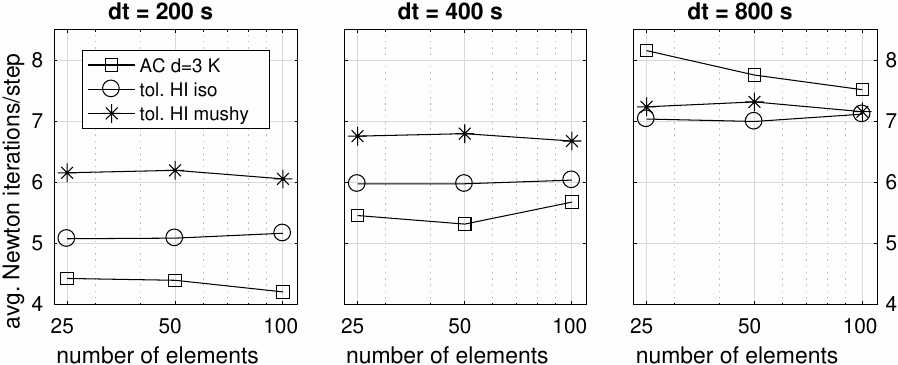}
    \caption{Melting volume example: average number of iterations of best performing methods for latent heat depending on spatial and temporal discretization.}
    \label{fig:compare_iterations_discretization_meltvol_reduced}
    \end{figure}

Next we turn to the efficiency of all methods and examine the average number of Newton iterations per time step shown in Fig.~\ref{fig:compare_iterations_discretization_meltvol}. Again, the original variants of the HI scheme need the most Newton iterations and are especially sensitive to temporal and spatial discretization. They are no longer considered in the remaining examinations. Fig.~\ref{fig:compare_iterations_discretization_meltvol_reduced} only compares iterations of the AC and tolerance-based HI methods. The iteration count increases with increasing time step size for all three methods but stays more or less constant over all spatial discretizations. In the rightmost graph showing the largest time step, the AC for the first time requires more iterations than the proposed tolerance-based HI methods.

\begin{figure}
    \centering
    \includegraphics[width=.8\linewidth]{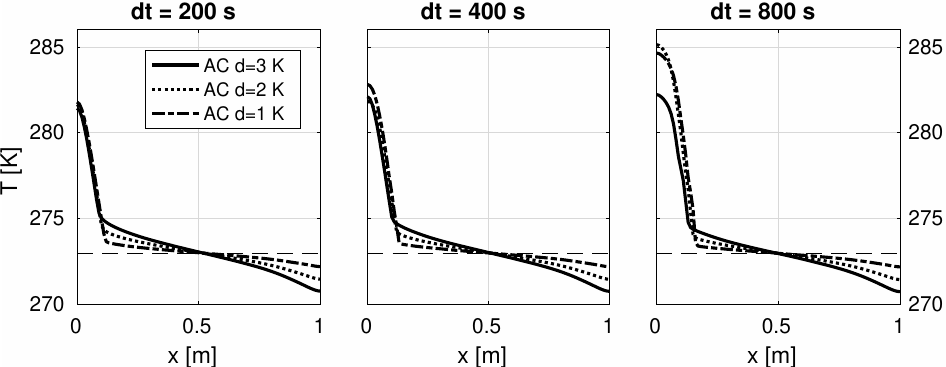}
    \caption{Melting volume example: temperature profiles at $t_f$ obtained with AC methods for three time step sizes and with different phase change interval widths $d$. 100 elements.}
    \label{fig:meltvol_appcapa_width_temp}
    \end{figure}

This melting volume example reveals an already mentioned problem of AC methods, namely the possibility to neglect much of latent heat by stepping over or passing through the phase change interval too fast. The AC method is used with three widths $d\in\{\unit[1]{K},\unit[2]{K},\unit[3]{K}\}$ to compute artificial solidus and liquidus temperatures $T_s = T_m -d$ and $T_l=T_m+d$. The final temperature profiles are graphed in Fig.~\ref{fig:meltvol_appcapa_width_temp} for three time step sizes and the finest mesh with 100 finite elements. Obviously, the profiles differ in the respective phase change intervals which would not be the main concern in PBFAM. Instead focus lies on the maximum temperature predicted on the left edge. For the smallest time step $\Delta t = \unit[200]{s}$ all widths reach almost the same maximum temperature. Increasing the step size leads to larger discrepancies in the maximum temperature. A smaller width $d$ correlates with higher maximum temperatures which in turn implies that not all latent heat has been absorbed.

\paragraph{Summary of preliminary simulations:} Taking into account all experience gathered from the two numerical examples, the authors recommend the use of the proposed tolerance-based HI method or an AC method. We found that the new criterion \eqref{eq:heatint_skip_tolerance} to determine nodes that undergo phase change is superior to the one originally introduced by \cite{Rolph1982}. This is due to two reasons. First, the accuracy is user-controllable by setting a respective tolerance. Second, the new stopping criterion~\eqref{eq:heatint_skip_tolerance} typically leads to a significant reduction of nonlinear Newton iterations except for scenarios with very strict tolerances, which are, however, not expected to be necessary in the targeted application PBFAM.

Naturally, it is hard to predict a-priori which method will lead to less Newton iterations since this depends on the specific problem, tolerances and the (artificial) phase change interval width. Therefore, in the following, an actual example in the context of PBFAM will be investigated.

\subsection{Single track scan}
\begin{figure}
    \centering
    \includegraphics[width=.6\linewidth]{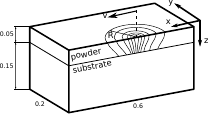}
    \caption{Geometry and moving laser heat source for single track scan example (all dimensions given in \unit{mm})~\cite{Hodge2014}.}
    \label{fig:single_track_scan_setup}
\end{figure}

The following example simulates the scanning of a single track in a PBFAM process and was introduced in \cite{Gusarov2009} and has also been simulated elsewhere, see \cite{Hodge2014}. A schematic sketch of the setup is shown in Fig.~\ref{fig:single_track_scan_setup}. A volumetric heat source described by Eq.~\eqref{eq:mm_gusarov_laser_model} with effective power $W=\unit[30]{W}$ and size $R=\unit[0.06]{mm}$ is applied to the powder domain.

The geometry consists of a cuboid of $0.6\times 0.2\times 0.2\, \unit{mm^3}$. The top layer of \unit[0.05]{mm} in $z$-direction is in powder form and rests on top of a solid substrate domain. The material is a 316L type steel with parameters summarized in Tab.~\ref{tab:hodge_params}. The whole domain is initialized at a fixed temperature $T_0=\unit[303]{K}$. All surfaces are insulating, only $x=\unit[0.6]{mm}$ is subject to an essential boundary condition $\hat{T}=T_0=\unit[303]{K}$.

\begin{table}
    \centering
    \caption{Material parameters for single track scan example \cite{Hodge2014}.}
    \label{tab:hodge_params}
    \begin{tabular}{llll}
    \toprule
    Parameter & Description & Value & Unit\\
    \midrule
    $\rho_h$ & Hemispherical reflectivity & 0.7 & -\\
    $\beta_h$ & Extinction coefficient & 60 &mm$^{-1}$\\
    $c_p$ & Heat capacity, powder & 2.98 & \unitfrac{MJ}{m$^3$K}\\
    $c_s$ & Heat capacity, solid & 4.25 & \unitfrac{MJ}{m$^3$K}\\
    $c_m$ & Heat capacity, melt & 5.95 & \unitfrac{MJ}{m$^3$K}\\
    $k_p(T)$ & Conductivity, powder & 0.2@200,&\\& & 0.3@1600 & \unitfrac{W}{mK}@K\\
    $k_c$ & Conductivity, solid/melt & 20 & \unitfrac{W}{mK}\\
    $T_m$ & Melting temperature & 1700 & K\\
    $h_m$ & Latent heat of fusion & 2.18 & \unitfrac{GJ}{m$^3$}\\
    \bottomrule
    \end{tabular}
    \end{table}

The laser beam center moves from an initial position at $x=-\unit[0.06]{mm}$ (one laser beam radius outside the domain) with a scanning speed of $v=\unitfrac[120]{mm}{s}$ in $x$-direction along the symmetry plane $y=0$.
The powder layer is discretized by a regular hexahedral mesh with element size $h_0= \unit[0.0025]{mm}$, i.e. $n_\text{ele}^z=20$ elements across the powder layer height. In the substrate domain, a mesh with double element height in $z$-direction is applied. Moreover, an adaptive time stepping scheme is applied that halves the time step size if no convergence is achieved by the employed Newton-Raphson scheme (within a prescribed maximal number of iterations), and that doubles the step size again after four convergent time steps on the smaller step size level. As initial step size a value of $\Delta t^{(0)} = \unit[1]{\mu s}$ has been chosen, which has been verified to yield a sufficiently small time discretization error.

\begin{figure}
    \centering
    \includegraphics[width=.9\linewidth]{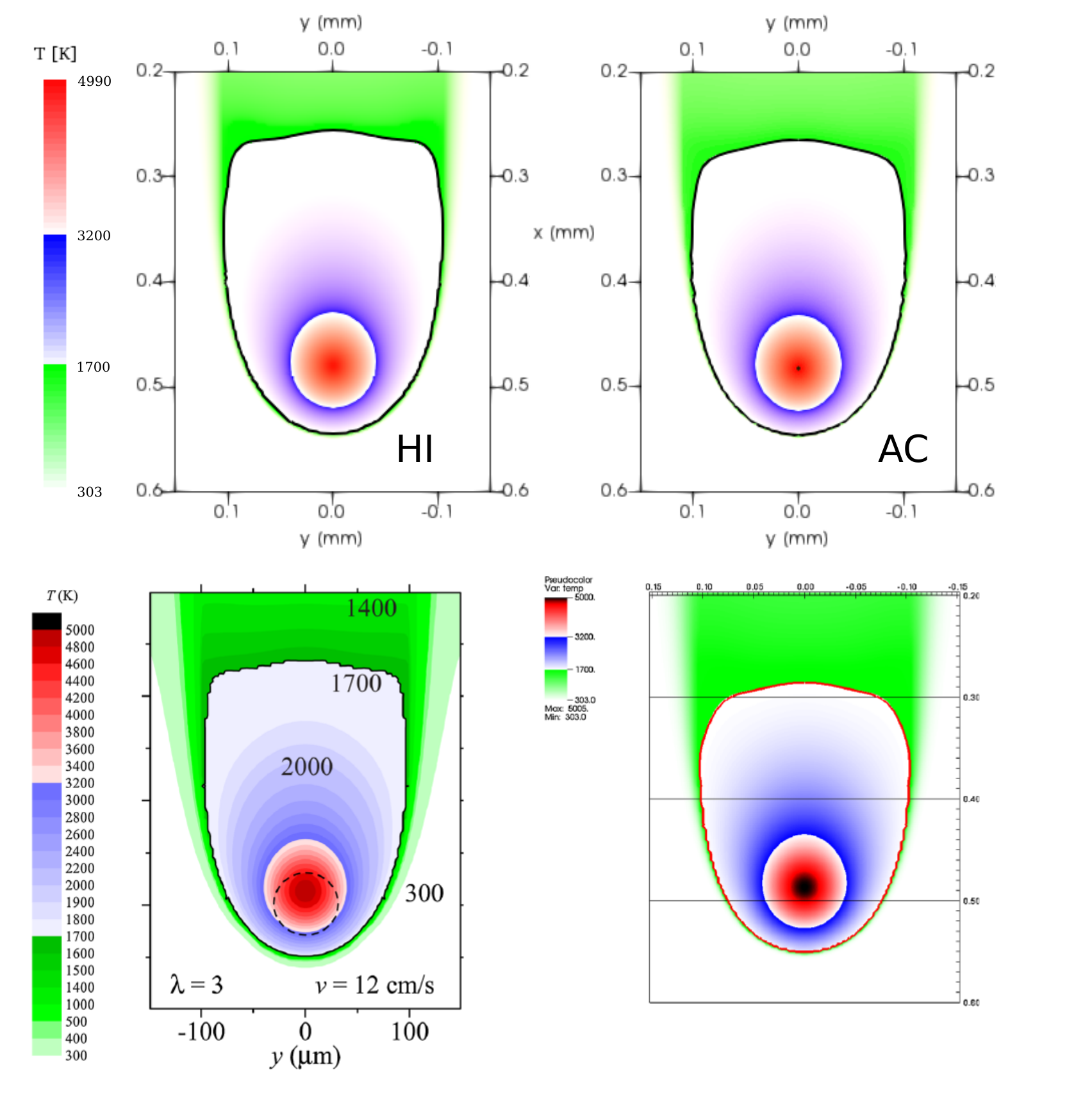}
    \caption{Surface temperature profile and melt pool shape in steady-state.
    Current results from HI scheme (top, left) and AC scheme (top, right) as well as
    reference results from Gusarov et al.~\cite{Gusarov2009} (bottom, left) and  Hodge et al.~\cite{Hodge2014} (bottom,right)}
    \label{fig:results_hodge_BACI_meltpool}
    \end{figure}

In prior simulations of this example, latent heat effects have been taken into account via an enthalpy method \cite{Gusarov2009} and an isothermal HI method \cite{Hodge2014}. Here, we will use an AC method and subsequently the newly proposed tolerance-based isothermal HI scheme to simulate the process.
An artificial melting interval is introduced for the AC method. As a baseline we chose $T_s = \unit[1600]{K}$ and $T_l = \unit[1800]{K}$, i.e., $d=\unit[100]{K}$. Isothermal HI is applied with a tolerance of $\varepsilon_\text{tol}=0.001$. 
The isothermal HI scheme is used in combination with enthalpy-based parameter interpolation, while the AC method is used with temperature-based interpolation.
In a first step, qualitative characteristics of the solution shall be discussed. After a short period of time the melt pool shape reaches a steady-state. Its geometry can be visualized by the isotherm $T=T_m$. Fig.~\ref{fig:results_hodge_BACI_meltpool} compares the results obtained with the AC and tolerance-based HI method to the results reported in~\cite{Gusarov2009} and~\cite{Hodge2014}. Both the AC and HI solution show good agreement with the reference. The melt pool dimensions and peak temperatures are compared quantitatively to the reference solutions in Tab.~\ref{tab:results_comparison_meltpool}. All compared quantities show good agreement.

    \begin{table}
        \centering
        \caption{Comparison of maximum temperature and melt pool dimensions resulting from the different latent heat models.}
        \label{tab:results_comparison_meltpool}
        \begin{tabular}{lllll}
        \toprule
        Quantity & \cite{Gusarov2009} & \cite{Hodge2014} & AC & HI\\
        \midrule
        Max. temp. [K] & 4900 & 5000 & 4990 & 4980\\
        Length [mm] & 0.30 & 0.27 & 0.28 & 0.29\\
        Width [mm] & 0.20 & 0.21 & 0.20 & 0.21\\
        Depth [mm] & 0.07 & 0.07 & 0.07 & 0.07\\
        \bottomrule
        \end{tabular}
        \end{table}

\begin{figure}
    \centering
    \includegraphics[width=.9\linewidth]{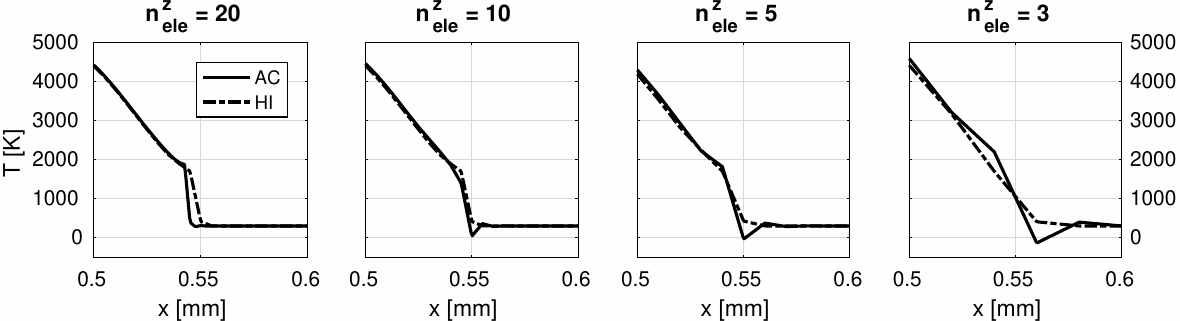}
    \caption{Surface temperature profiles for AC and HI along laser path (i.e. $y=\unit[0.0]{mm}$) for different spatial discretizations. Zoomed in to melting front of melt pool.}
    \label{fig:results_xcut_front}
    \end{figure}

Since no more quantitative data is provided by the reference papers, we compare AC and HI method with each other. In the preliminary examples in Section~\ref{sec:results_solidifcation_front} and \ref{sec:results_melting_vol} a strong dependency on spatial discretization was recognized. Three additional, coarser spatial discretizations with elements of size $2h_0$, $4h_0$ and  $\frac{20}{3}h_0$ (which results in $n_\text{ele}^z \in \{10,5,3\}$ elements over the powder layer height) are introduced to investigate this phenomenon for the single track scan. The accuracy of both methods can be judged by looking at characteristic temperature profiles in the steady-state. All results presented in the following are shown for (approximately) $t=\unit[4.6]{\mu s}$, which is the same point in time for which the melt pool shape has been illustrated in Fig.~\ref{fig:results_hodge_BACI_meltpool}.  First, Fig.~\ref{fig:results_xcut_front} shows the surface temperature profiles for all discretizations plotted along the laser path (i.e. $y=0$) in the vicinity of the melting front. This front is characterized by high temperature gradients and rates. With increasing element size, larger nonphysical oscillations in the temperature profile are observed for the AC scheme. These oscillations can be traced back to the limitation of the employed first-order finite elements in representing strong gradients and material nonlinearities, here mainly caused by the extreme gradients of the thermal conductivity across the phase interface. Employing finite elements based on higher order shape functions can remedy this numerical issue~\cite{OHara2011,Kollmannsberger2018}.

No such oscillations occur with the HI scheme, which enforces the temperature in the phase transition region to lie within a temperature interval (implicitly) prescribed through the tolerance $\varepsilon_\text{tol}$. Although the shape functions used with the HI scheme are still first-order, the reset of temperature to a consistent melt temperature as described in \eqref{eq:mm_heatint_tempreset} seems to prohibit temperature oscillations in the phase interface region as observed for AC, even though the HI scheme performs phase change and parameter interpolation within a considerably smaller temperature interval.
It is important to note that so far the oscillations have not been observed to cause stability issues (e.g. a significant energy increase in the discrete system) and they remain small compared to the overall temperature range.

\begin{figure}
    \centering
    \includegraphics[width=.9\linewidth]{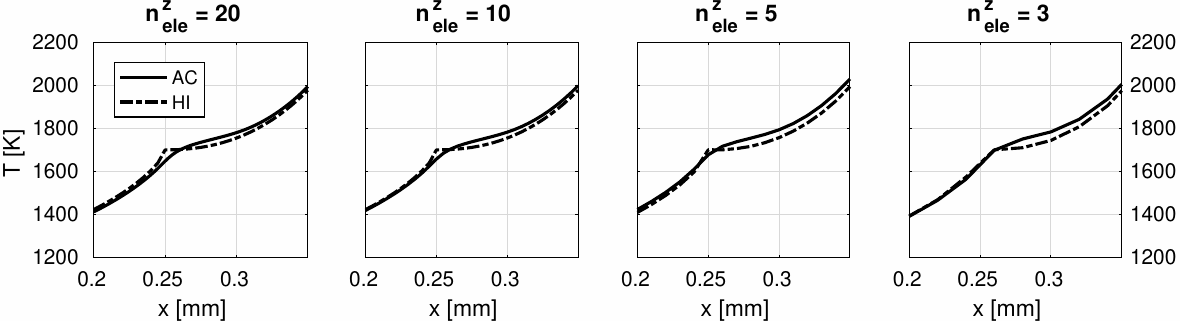}
    \caption{Surface temperature profiles for AC and HI along laser path (i.e. $y=\unit[0.0]{mm}$) for different spatial discretizations. Zoomed in to solidification region in melt pool tail.}
    \label{fig:results_xcut_back}
    \end{figure}

A second phase change happens along the laser path (i.e. $y=\unit[0.0]{mm}$) when material cools down again at the tail of the melt pool. Fig.~\ref{fig:results_xcut_back} gives a detailed view of the temperature profile in this region. 
The HI method produces a kink in the temperature profile at $T_m$, which is to be expected for a phase interface.
The AC method produces a mushy phase change region and no kink is observed. However, further away both temperature profiles are in good agreement again.

\begin{figure}
    \centering
    \includegraphics[width=.9\linewidth]{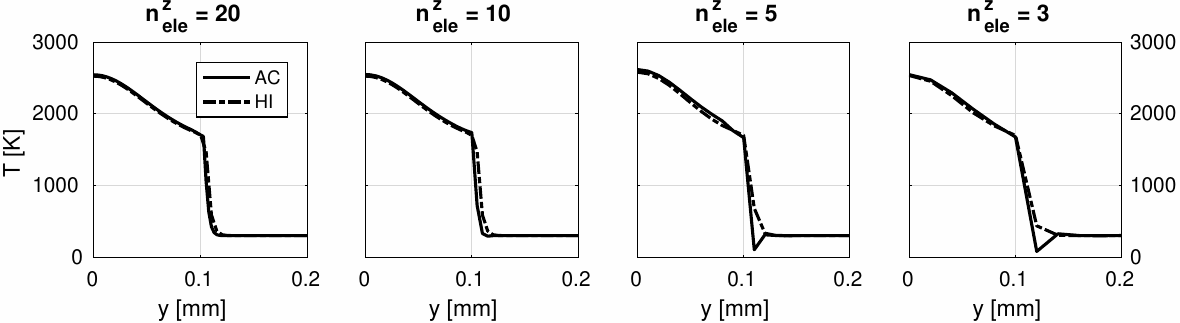}
    \caption{Surface temperature profiles for AC and HI transverse to laser path at $x=\unit[0.4]{mm}$ showing transition from melt pool to powder.}
    \label{fig:results_ycut_melt}
    \end{figure}

\begin{figure}
    \centering
    \includegraphics[width=.9\linewidth]{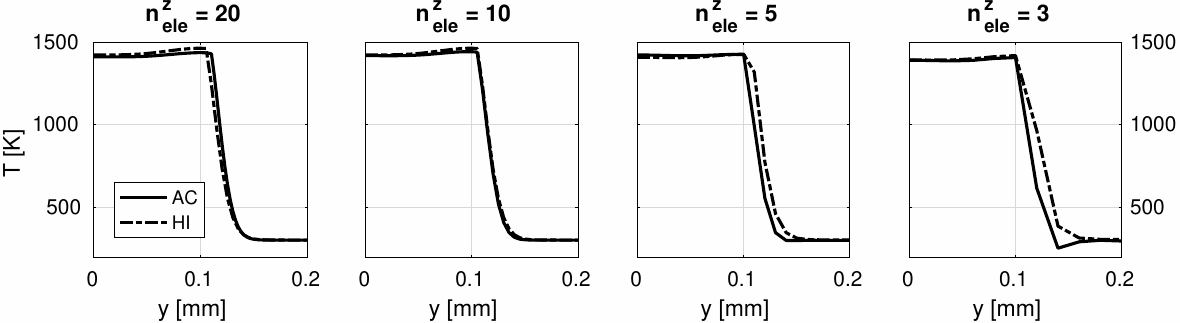}
    \caption{Surface temperature profiles for AC and HI transverse to laser path at $x=\unit[0.2]{mm}$ showing transition from solid to powder.}
    \label{fig:results_ycut_solid}
    \end{figure}

Another aspect to investigate is the resulting temperature profile transverse to the laser scanning direction. Change from melt to powder can be investigated with a cut in $y$-direction at $x=\unit[0.4]{mm}$ as shown in Fig.~\ref{fig:results_ycut_melt}. Again oscillations appear in the AC solution for coarser meshes but not in the HI solution. A similar picture emerges for a cut in $y$-direction through solid and powder, e.g. at $x=\unit[0.2]{mm}$ as illustrated in Fig.~\ref{fig:results_ycut_solid}. In both transverse cuts temperature profiles differ for AC and HI method in proximity to $T_m$ and agree well in some distance to the phase interface.

\begin{figure}
    \centering
    \includegraphics[width=.6\linewidth]{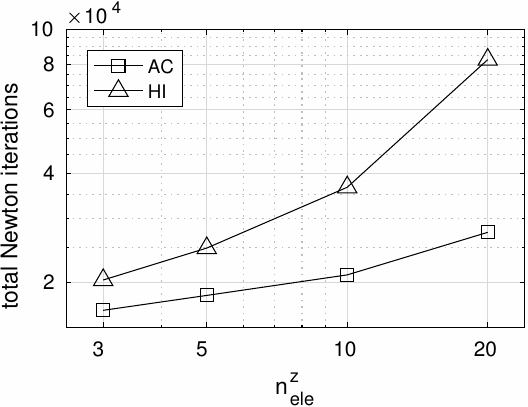}
    \caption{Numerical efficiency of AC and HI in terms of total Newton iterations required.}
    \label{fig:results_hodge_totaliter}
    \end{figure}

As suspected, in PBFAM applications the chosen method for latent heat only has a very local influence on the resulting temperature field, but does not significantly affect the global temperature characteristics from a rather macroscopic point of view. Another aspect of importance for PBFAM process simulations is numerical efficiency, here assessed in terms of nonlinear solver performance. Fig.~\ref{fig:results_hodge_totaliter} depicts the total number of Newton iterations accumulated over the whole simulation time for the different spatial discretizations. The results of the preliminary numerical examples in the previous sections seem to transfer to a larger example: The HI method shows a strong dependency on the spatial resolution. The difference in the number of Newton iterations between AC and HI method is less pronounced  in the practically relevant range of discretizations (e.g. 3 elements across the powder layer thickness).

\begin{figure}
    \centering
    \includegraphics[width=.6\linewidth]{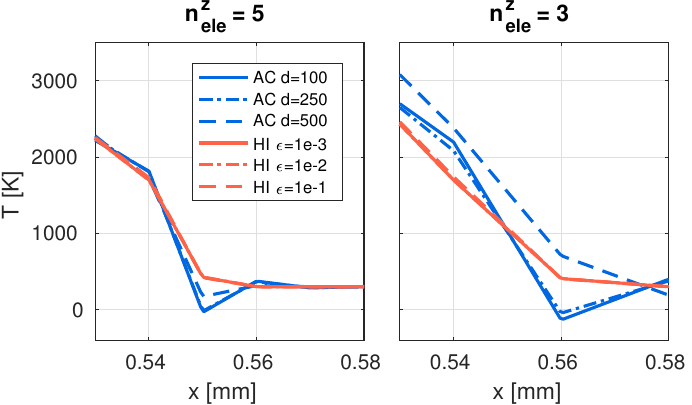}
    \caption{Surface temperature profiles for AC with different phase change intervals along laser path (i.e. $y=\unit[0.0]{mm}$). Zoomed in to melting front.}
    \label{fig:hodge_xcut_front_appcapa_d}
    \end{figure}

\begin{figure}
    \centering
    \includegraphics[width=.6\linewidth]{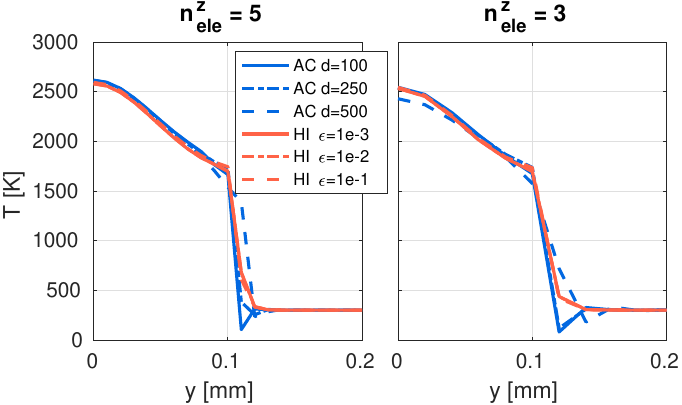}
    \caption{Surface temperature profiles for AC with different phase change intervals transverse to laser path at $x=\unit[0.4]{mm}$ showing transition from melt pool to powder.}
    \label{fig:hodge_ycut_melt_appcapa_d}
    \end{figure}

\begin{figure}
        \centering
        \includegraphics[width=.6\linewidth]{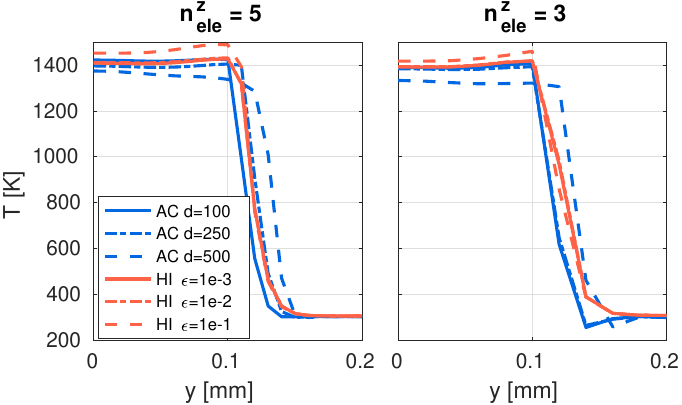}
        \caption{Surface temperature profiles for AC with different phase change intervals transverse to laser path at $x=\unit[0.2]{mm}$ showing transition from solid to powder.}
        \label{fig:hodge_ycut_solid_appcapa_d}
        \end{figure}

In a next step, it shall be investigated how the accuracy and numerical efficiency of the two considered phase change schemes can be influenced by the respective numerical parameters $d$ (phase change interval of the AC scheme) and $\varepsilon_\text{tol}$ (tolerance of the HI scheme). In a first step, the phase change interval for the AC method is varied for the two coarsest spatial discretizations to see how it affects the temperature oscillation. It has to be noted that this also changes the temperature interval for temperature-based parameter interpolation. The intervals are given by $T_s = T_m - d$ and $T_l = T_m +d$, where $d \in \{100,250,500\} \unit{[K]}$. Moreover, two additional versions of HI with relaxed tolerances of $\varepsilon_\text{tol}=0.01$  and $\varepsilon_\text{tol}=0.1$ are simulated. The resulting surface temperatures are plotted in Fig.~\ref{fig:hodge_xcut_front_appcapa_d}, \ref{fig:hodge_ycut_melt_appcapa_d} and \ref{fig:hodge_ycut_solid_appcapa_d} in the regions that showed oscillations in the previous plots. While the increased phase change interval for the AC decreases the amplitude of these oscillations by a certain amount, the overall accuracy of the temperature profiles decreases as well. Thus, it has to be concluded that for a given spatial discretization the width $d$ of the phase change interval is not a suitable parameter to control the accuracy of AC schemes. The solutions obtained with different tolerances for the HI method show no oscillation. The solution from HI with tolerance $\varepsilon_\text{tol}=0.01$ is indistinguishable from the one with stricter tolerance $\varepsilon_\text{tol}=0.001$. The solution from HI with tolerance $\varepsilon_\text{tol}=0.1$ deviates slightly from the ones with stricter tolerances in the solid-powder transition region, see Fig.~\ref{fig:hodge_ycut_solid_appcapa_d}. However, it still seems to be considerably more accurate than the solution from AC with $d=\unit[500]{K}$.

\begin{figure}
    \centering
    \includegraphics[width=.6\linewidth]{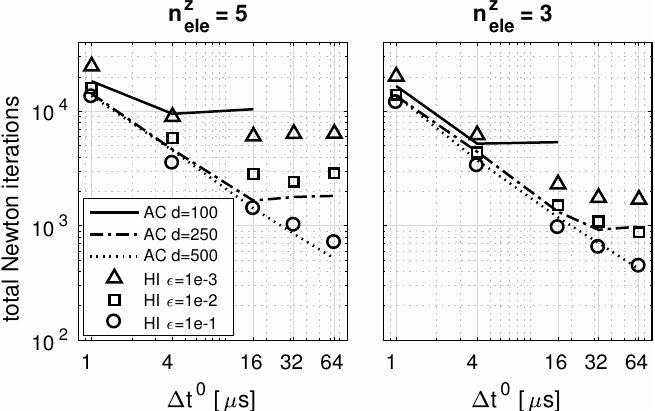}
    \caption{Total number of Newton iterations required for different initial time step sizes $\Delta t^0$ in the time step halving scheme.}
    \label{fig:hodge_timeadapt_iterations}
\end{figure}

Finally, the initial time step size $\Delta t^0$ for the three considered HI and AC variants is increased up to 64 times of the original value. The total number of Newton iterations that is now required is depicted  in~Fig.~\ref{fig:hodge_timeadapt_iterations}. It is emphasized that the time step halving scheme is still employed. It was optimized individually for each method to yield low iteration counts. Moreover, it has been checked that the time discretization error is still sufficiently small, i.e. there is no visible difference in the results for base time step sizes up to 16 times $\Delta t^0$ for all variants. Higher step sizes lead to visible albeit small deviations in the temperature profiles, which are smaller, however, than the deviations resulting from the different HI and AC variants. According to~Fig.~\ref{fig:hodge_timeadapt_iterations}, an increased phase change interval $d$ in AC allows for larger step sizes and thus less iterations. Especially on the coarsest mesh, HI with a high tolerance ($\varepsilon_\text{tol}=0.1$) yields a comparable number of accumulated Newton iterations, but at a considerably increased accuracy, as compared to the AC method with the (unphysically) large phase change interval $d=\unit[500]{K}$. All in all, for a given spatial discretization, the proposed HI scheme allows to directly control numerical efficiency and accuracy by means of a user-defined tolerance. Within the considered scope of numerical examples and practically relevant spatial and temporal discretizations, this property made the novel tolerance-based HI scheme preferable as compared to the original HI scheme and the investigated AC method.

\section{Conclusion}
\label{sec:conclusion}
In this work, an extension of phase change and latent heat models for the simulation of metal powder bed fusion additive manufacturing processes on the macroscale has been proposed and different models have been compared with respect to accuracy and
numerical efficiency. In this context, a systematic formulation of phase fraction variables has been proposed relying either on temperature- or enthalpy-based interpolation schemes. Moreover, latent heat has been considered either by means of an apparent capacity (AC) or heat integration (HI) method. For the latter, a novel phase change criterion has been proposed, which combines superior accuracy and numerical efficiency (in terms of an improved  nonlinear solver performance allowing for larger time step sizes and fewer iterations per time step) as compared to the original HI scheme. Compared to the AC approach, the numerical efficiency of the proposed tolerance-based HI scheme is comparable while offering an increased level of accuracy.
Numerical results from the literature have been reproduced, which shows the validity of the proposed scheme. In summary, both the AC and the proposed tolerance-based HI scheme perform well when considering the accuracy requirements as well as practically relevant spatial and temporal discretization resolutions for PBFAM process simulation. Specifically, global temperature characteristics, such as the peak temperature, can be accurately captured with both methods. Still, the authors believe that the new tolerance-based HI method is advantageous over AC schemes due to the user-controllable tolerance, which allows to directly control numerical efficiency and accuracy of the scheme, and which can directly be interpreted as the error in latent heat made during a phase change process.

For part-scale PBFAM models thermo-mechanical interaction is one of the primary interests. Structural parameters such as Young's modulus or the thermal expansion coefficient may depend upon the temperature history in a similar fashion as proposed in the current work for the thermal parameters. Future research work will focus on the question of how the proposed methods for latent heat and parameter interpolation will behave in the thermo-mechanically coupled scenario.

% \section*{Availability of data and materials}
% The research code, numerical results and
% digital data obtained in this project are held on deployed servers that are backed up daily. Fundamental
% work results are also stored on a NAS server managed by the Leibniz Rechenzentrum (LRZ) in Garching. The datasets used and/or analysed during the current study are available from the corresponding author on reasonable request.

% \section*{Competing interests}
%  The authors declare that they have no competing interests.
  
% \section*{Funding}
%  Not applicable.

% \section*{Author's contributions}
%  All authors contributed to the derivation of model equations, the discussion of results, and writing the manuscript. In addition, SP conducted the specific code implementation and the shown numerical simulations. CM and WAW worked out the general conception of the proposed modeling approach. All authors read and approved the final manuscript.

\section*{Acknowledgments}
This work was supported by the German Research Foundation (DFG) and the Technical University of Munich (TUM) in the framework of the Open Access Publishing Program.

%%  After submission of the .TEX file,                     %%
%%  you will be prompted to submit your .BBL file.         %%

% if your bibliography is in bibtex format, use those commands:
\bibliographystyle{vancouver} % Style BST file (bmc-mathphys, vancouver, spbasic).
\bibliography{ref}      % Bibliography file (usually '*.bib' )

%%%%%%%%%%%%%%%%%%%%%%%%%%%%%%%%%%%
%%                               %%
%% Additional Files              %%
%%                               %%
%%%%%%%%%%%%%%%%%%%%%%%%%%%%%%%%%%%

% \section*{Additional Files}
%   \subsection*{Additional file 1 --- Sample additional file title}
%     Additional file descriptions text (including details of how to
%     view the file, if it is in a non-standard format or the file extension).  This might
%     refer to a multi-page table or a figure.

%   \subsection*{Additional file 2 --- Sample additional file title}
%     Additional file descriptions text.

\end{document}